\documentclass{article}
\usepackage{arxiv}
\pdfoutput=1
\usepackage[utf8]{inputenc} 
\usepackage[T1]{fontenc}    
\usepackage{hyperref}       
\usepackage{url}            
\usepackage{booktabs}       
\usepackage{amsfonts}       
\usepackage{nicefrac}       
\usepackage{microtype}      
\usepackage{lipsum}		
\usepackage{graphicx}
\usepackage{doi}
\usepackage{amsfonts}
\usepackage{amsmath}
\usepackage{xcolor}
\usepackage{bm}
\usepackage{amssymb}

\title{Intra-community link formation and modularity in ultracold growing hyperbolic networks}

\author{ \href{https://orcid.org/0000-0003-3417-2655}{\includegraphics[scale=0.03]{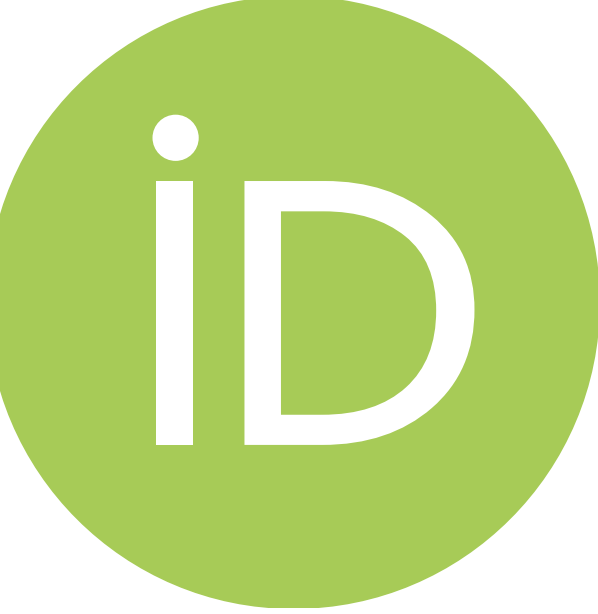}\hspace{1mm}Sámuel G. Balogh}
        \\
	National Laboratory for Health Security, Alfréd Rényi Institute of Mathematics\\
	Re{\'a}ltanoda utca 13-15, Budapest, 1053, Hungary \\
 Department of Biological Physics, Eötvös Lor{\'a}nd University\\
 P{\'a}zm{\'a}ny P.\ stny.\ 1/A, Budapest, H-1117, Hungary\\
	\texttt{balogh@hal.elte.hu} \\
	\And
	\href{https://orcid.org/0000-0002-3406-4200}{\includegraphics[scale=0.03]{orcid.png}\hspace{1mm}Gergely Palla} \\
	Department of Biological Physics, Eötvös Lor{\'a}nd University\\
        P{\'a}zm{\'a}ny P.\ stny.\ 1/A, Budapest, H-1117, Hungary\\
        Health Services Management Training Centre, Semmelweis University\\ 
        Kútvölgyi út 2., Budapest, H-1125, Hungary
}

\renewcommand{\shorttitle}{\textit{arXiv} Template}

\hypersetup{
pdftitle={Intra-community link formation and modularity in ultracold growing hyperbolic networks},
pdfauthor={},
pdfkeywords={},
}

\begin{document}
\maketitle

\begin{abstract}
Hyperbolic network models, centered around the idea of placing nodes at random in a hyperbolic space and drawing links according to a probability that decreases as a function of the distance, provide a simple, yet also very capable framework for grasping the small-world, scale-free, highly clustered and modular nature of complex systems that are often referred to as real-world networks. In the present work we study the community structure of networks generated by the Popularity Similarity Optimization model (corresponding to one of the fundamental, widely known hyperbolic models) when the temperature parameter (responsible for tuning the clustering coefficient) is set to the limiting value of zero. By focusing on the intra-community link formation we derive analytical expressions for the expected modularity of a partitioning consisting of equally sized angular sectors in the native disk representation of the 2d hyperbolic space. Our formulas improve earlier results to a great extent, being able to estimate the average modularity (measured by numerical simulations) with high precision in a considerably larger range both in terms of the model parameters and also the relative size of the communities with respect to the entire network. These findings enhance our comprehension of how modules form in hyperbolic networks. The existence of these modules is somewhat unexpected, given the absence of explicit community formation steps in the model definition.

\end{abstract}


\tableofcontents

\section{Introduction}
In the past couple of years, a vast number of studies have pointed out that the patterns of interactions between the fundamental units of various complex systems can be effectively described using network theory~\cite{Dorog_book,Newman_Barabasi_Watts,Jari_Holme_Phys_Rep,Vespignani_book}. Since then, this interdisciplinary field has gained widespread popularity and shown that real-world networks can demonstrate an immensely rich behavior~\cite{Laci_revmod, boccaletti_rev}. Despite this phenomenological richness, the inhomogeneous scale-free nature of the degree distribution~\cite{Laci_science}, the relatively large average clustering coefficient~\cite{Watts-Strogatz}, the small-world property~\cite{Milgram_small_world}, and the hierarchical organisation~\cite{zafeiris_Vicsek_hier,Sole_hier_soc,mesh} are all important features of complex networks that show universality across a broad spectrum of systems ranging from the level of metabolic networks within cells to the level of the entire society~\cite{Laci_revmod,Dorog_book}. Moreover, the vast majority of real networks also exhibit a pronounced community structure~\cite{Fortunato_coms,Fortunato_Hric_coms}, that is, they contain locally denser sub-graphs at the mesoscopic organizational scale. 

Besides exploring the occurrence of networks in a wide variety of complex systems, another key goal of network theory has always been to find simple mathematical models that can account for the above universal features~\cite{Laci_science, Watts-Strogatz, PSO, caldarelli_general_sf_recipe,gen_treshold_scirep,triadic_closure_holme}. A remarkable and very elegant approach in this direction is given by hyperbolic models~\cite{hyperGeomBasics,S1,PSO,dPSO}, in which the nodes of the networks are placed randomly into a hyperbolic space of a given dimension $d$, and the connections between them are formed based on a probability decreasing with their hyperbolic distance. In general, hyperbolic network models can be viewed as a particular case of a broader hidden variable framework~\cite{caldarelli_first_peaked,caldarelli_general_sf_recipe,boguna_general_hv,Garlaschelli_Entropy,gen_treshold_scirep}, where the hidden variables of the nodes are associated with the coordinates of the nodes in the hyperbolic space, whereas the connection probability between pair of nodes depends specifically on their respective distances.

One of the most well-studied model of the above kind is given by the popularity-similarity optimisation (PSO) model~\cite{PSO}, which simulates the dynamic evolution of networks over the two-dimensional hyperbolic space $\mathbb{H}^2$. In this model, nodes are incrementally added to the native disk representation of $\mathbb{H}^2$~\cite{hyperGeomBasics}, until the network accumulates a total number of $N$ nodes. At each time step 
$t=1,\dots, N$, the new node $t$ joining the network is assigned an initial radial coordinate $r_{t}(t)=r_t$ that depends logarithmically on $t$, and also an angular coordinate chosen uniformly at random. During the growth of the network, angular coordinates remain unaltered, however, the nodes gradually move outwards from the center of the disk in a radial direction. More precisely, at each timestep $t$ the radial coordinate of each previously (at time $s<t$) appeared node $s$ is increased according to
\begin{equation}
    r_{s}(t)=\beta r_{s}+(1-\beta)r_{t},
\end{equation}
where $\beta$ referred to as the popularity fading parameter that characterizes the rate of the nodes' outward drift~\cite{PSO}. Once a new node $t$ appears, it establishes connections with, on average, $m$ number of previously added nodes. The probability for node $t$ to get connected to a specific already-existing node $s<t$ depends on their hyperbolic distance $x_{st}$ as 
\begin{equation} p(x_{st})=\frac{1}{1+e^{\frac{\zeta}{2T}(x_{st}-R_t)}},
\label{eq:PSO_con_prob}
\end{equation}
where $\zeta = \sqrt{-K}$ parametrises the curvature of the hyperbolic space, $T\geq 0$ is a temperature-like model parameter, and $R_t$ is the so-called cutoff distance (for a more detailed description of the PSO model see \ref{sec:appendixA}).

Very recently, extensive numerical investigations have unveiled a remarkable characteristic of networks generated by the PSO model. These networks inherently exhibit a prominent community structure across a wide range of parameters, despite the absence of any explicitly designed community formation rule within the model definition~\cite{commSector_commDetMethod,our_hyp_coms}. Subsequently, this observation was verified in Ref.~\cite{Balogh2023} by means of analytical arguments, wherein the authors showed that networks generated by the PSO model with temperature $T<1/2$ can become extremely modular in the thermodynamic limit, i.e. when $N\to \infty$. The present article delves deeper into the intricacies of this calculation, offering substantially improved predictions specifically tailored for the evaluation of modularity in ultracold ($T=0$) PSO networks. In addition to the previous advancements, we explore the establishment of intra-community links in this particular limit and its connection to a non-linear preferential attachment rule. Last, but not least we address the limiting scenario of Random Geometric Graphs~\cite{penrose_rgg} ($T=\beta=0$), examine the conditions under which ultracold PSO networks exhibit maximum modularity, and present various other intriguing results associated with them.

\section{Preliminaries: uniform partitioning scheme and the modularity}

Previous numerical studies on hyperbolic networks reported a remarkable observation regarding the modular structure of these networks; modules identified by state-of-the-art community finding methods inherently align with distinct angular regions on the hyperbolic disk, commonly referred to as angular sectors~\cite{commSector_commDetMethod,commSector_hypEmbBasedOnComms_2016,commSector_hypEmbBasedOnComms_2019,our_hyp_coms}. This alignment is consistent across various community detection methods, all of which operate without any prior knowledge of the underlying hyperbolic metric~\cite{our_hyp_coms}. 

The previous correspondence between angular regions of the hyperbolic plane and the network communities served as inspiration for a simplified, yet highly informative partitioning scheme outlined in Ref.~\cite{Balogh2023}, and also discussed in Refs.~\cite{commSector_commDetMethod, Mod_of_RHG}. In particular, Ref.~\cite{Balogh2023} defines the communities of PSO networks based on the angular coordinates of nodes,
resulting in the division of the native disk into $q$ number of communities, each possessing an identical angular width of $2\pi/q$. Here, we adopt the same strategy and assess modularity $Q$ as a quality metric to gauge how effectively this uniform partitioning compartmentalizes the network's communities~\cite{Newman_modularity_original}. Introducing $b_s$ as the number of intra-community links of node $s$, the modularity $Q$ is written as
\begin{equation}
Q=\sum\limits_{c=1}^{q}\left [ \frac{\sum\limits_{s\in c} b_s}{2E}-\left ( \frac{\sum\limits_{s \in c}k_{s}}{2E} \right )^2\right ],
\label{eq:Q_exact}
\end{equation}
where the summation goes over the communities, $k_s$ is the degree of the community member $s$, and $E$ stands for the total number of links in the network~\cite{Newman_modularity_original}. With the above simplifications in place, the expected modularity of PSO networks in the asymptotic limit is now expressed as a function of both the system size, denoted as $N$, and the free parameter $q$. This results in the following equation:
\begin{equation}
   \bar{Q}\left(N,q\right) \approx\sum\limits^N_{s=1} \frac{\bar{b}_{s}(N,q)}{2E}-\frac{1}{q},
    \label{eq:av_Q_base0}
\end{equation}
as demonstrated in the study by Balogh \textit{et al.}~\cite{Balogh2023} under the conditions of $q\ll N$ and $T<1/2$.
The above formulation relies on a type of homogeneous mixing approximation, which typically works best for homogeneous networks at $\beta \ll 1$ values, whereas becoming less accurate in the presence of hubs at $\beta \approx 1$. 

The upcoming sections convey two crucial messages: firstly, we reaffirm the validity of Eq.(\ref{eq:av_Q_base0}), specifically focusing on the $T=0$ case. Secondly, and perhaps more significantly, we present an analytical expression for $\bar{Q}(N,q)$ appearing in Eq.(\ref{eq:av_Q_base0}) at $T=0$, offering not only improved accuracy compared to the approximation introduced in~\cite{Balogh2023}, but also extended applicability at basically any arbitrary value of $q$. 
In the subsequent section, we proceed with an in-depth exploration of ultracold PSO networks and their key characteristics, along with an examination of the foundational properties governing the formation of intra-community links.

\section{Results}
\subsection{Intra-community link formation in the PSO model at $T=0$}

Drawing parallels from statistical physics, in PSO networks at $T=0$, only the shortest links exist, much like electrons occupying the lowest energy states available to them at absolute zero temperatures.  More specifically, the connection probability $p(x_{st})$ between any pair of nodes $s$ and $t$ is either 1 or 0, depending on whether their hyperbolic distance $x_{st}$ is greater or less than the connectivity cut-off $R_t$, which functions analogously to the Fermi energy in this comparison. 
To articulate this more rigorously, the probability of nodes $s$ and $t$ being connected in the limit $T\to 0$ is precisely defined by a reversed Heaviside function~\cite{PSO}, given by
\begin{equation}
    \lim_{T\to 0}p(x_{st}) = \Theta\left(\Delta\theta_{st}-\frac{2}{X(s,t)}\right) =\left\{ \begin{array}{lll}
1,\ \text{if} \ \Delta\theta_{st}\leq\frac{2}{X(s,t)}\\0, \ \text{otherwise.}
\end{array}\right. 
\label{eq:heaviside}
\end{equation}
where $\Theta$ denotes the reversed Heaviside step function, $X(s,t)=\frac{2}{\pi}s^{\beta}\frac{t^{1-\beta}-1}{m(1-\beta)}$ is a new variable taking into account the intrinsic rules of the PSO model for defining the radial coordinates of the nodes and $\Delta\theta_{st}$ is the angular distance between $s$ and $t$. For further details, see \ref{subsubsec:PSOT0connection}.

Even though the angular 
coordinate of the nodes is uniformly distributed across the entire hyperbolic disk~\cite{PSO}, this uniformity does not hold when considering the angular difference between nodes falling into the same angular sector, corresponding to one of the communities. 
 As outlined in Ref.~\cite{Balogh2023}, the probability density of $\Delta\theta_{st}$ specifically for nodes $s,t$, that both belong to the same angular sector (with a central angle of $2\pi/q$),  
can be expressed as
\begin{equation}
\varrho(\Delta\theta_{st})=\frac{q}{\pi}-\frac{1}{2}\left(\frac{q}{\pi}\right)^2\Delta\theta_{st}.
    \label{eq:intra_angular_diff}
\end{equation}
After combining Eqs.(\ref{eq:heaviside}-\ref{eq:intra_angular_diff}), the probability that node $s$ forms an intra-community link with a new-coming node $t$ (given that the entire hyperbolic disk is divided into $q$ number of equally sized angular sectors and that $t$ arrives to the same sector as $s$), can be written as 
\begin{equation}
    \Pi_q(s,t) =
    \frac{1}{q}\int\limits^{2\pi/q}_{0}\left(\frac{q}{\pi}-\frac{1}{2}\left(\frac{q}{\pi}\right)^2\Delta\theta_{st}\right)\Theta\left(\Delta\theta_{st}-\frac{2}{X(s,t)}\right)\mathrm{d}\Delta\theta_{st}.
    \label{eq:piq}
\end{equation} 
The key point here is to recognize that the behavior of $\Pi_q$ is completely controlled by the argument value of the Heaviside function $\Theta$ in Eq.(\ref{eq:piq}). 
More specifically, Eq.(\ref{eq:piq}) can be rewritten in a simplified form yielding
\begin{align}
             \Pi_q(s,t) = \left\{ \begin{array}{lll}
\frac{2}{\pi}\frac{1}{ X(s,t)}-\frac{q}{\pi^2}\frac{1}{X^2(s,t)},\ \ \  \text{if} \ \ \ \frac{2}{X(s,t)}< \frac{2\pi}{q},\\
\\
q^{-1},\ \ \  \text{if} \ \ \ \frac{2}{X(s,t)}\geq \frac{2\pi}{q}.
\end{array}\right.
\label{eq:pi_cases}
\end{align}
the detailed derivation of which is moved to \ref{subsec:linkformationprob}.

\begin{figure}[htb]
\includegraphics[width=1.\textwidth]{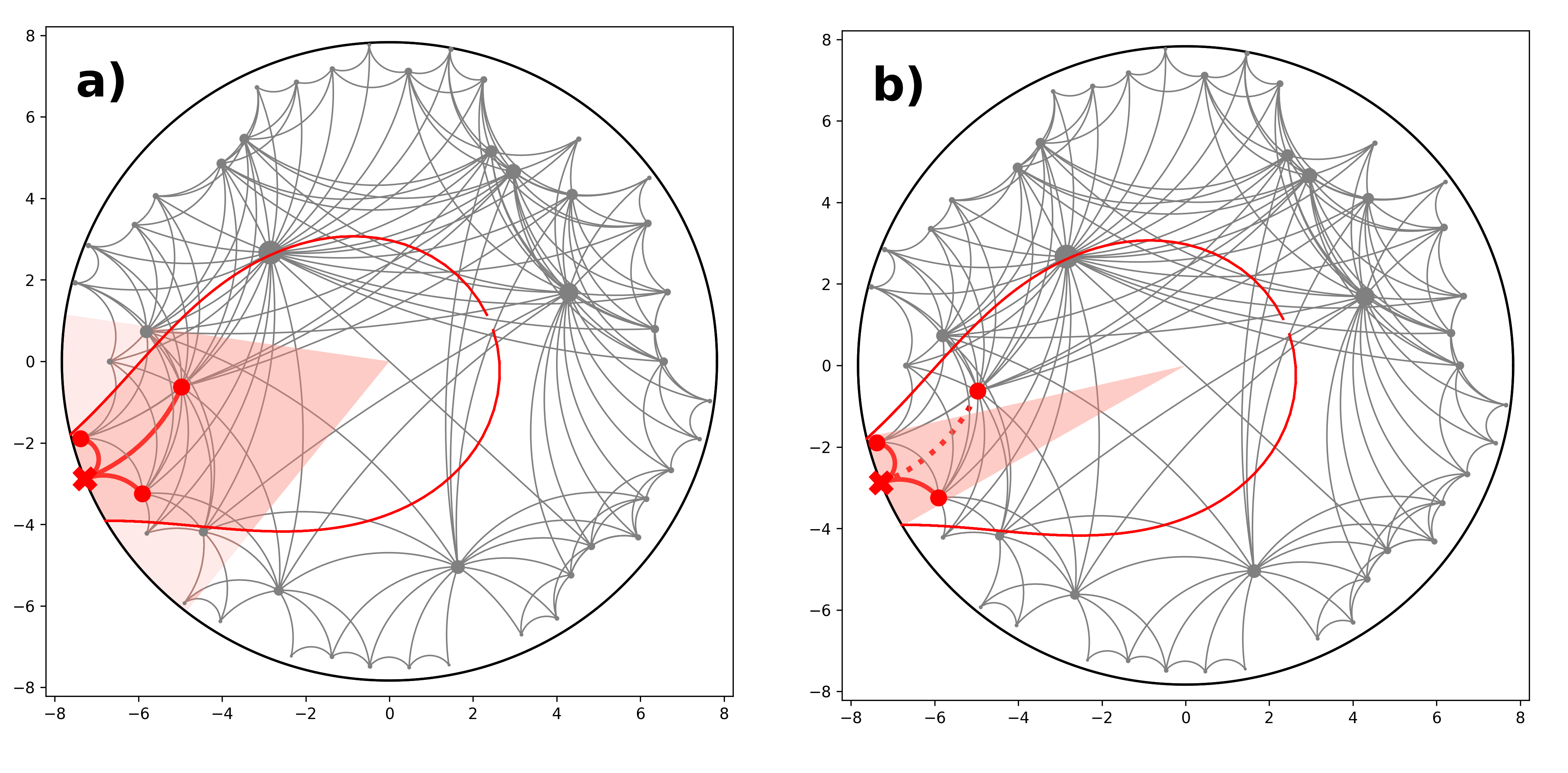}
\caption{\textbf{ The relationship between the number of intra-community links of a newly added node and the scope of the angular sector associated with its community in a given PSO network generated at $T=0$.
} 
In each of the two panels, the identical newly introduced node $t$ is denoted by a red cross symbol, along with red circles indicating nodes within a disk of radius $R_t$ around this node (visualized by a red drop-like curve). a) Illustrates a scenario where the angular scope of 
node $t$'s community (marked by the lighter shaded region) exceeds the angular connectivity range, leading to the formation of intra-community links (shown by solid red lines) only within the darker shaded region. In panel b), the opposite situation is shown, with the community's angular 
scope being small, resulting in intra-community links between node $t$ and every other node within the shaded angular sector. Both panels depict the same PSO network generated under model parameters $N=50, T=0, \beta=0.5, m = 4$.
\label{fig:intra_connection_ranges_illustration}}
\end{figure}

The interpretation of the previous result is the following: for each pair of nodes $s,t$ we can define two types of characteristic angular distances. The first one, expressed as $\varphi^{\text{c}}_{st}:=2/X(s,t)$, signifies an effective angular connectivity range between nodes $s$ and $t$, which can be derived from the cut-off value of the connection function $R_t$ appearing in Eq.(\ref{eq:PSO_con_prob}). More specifically, it indicates that when the actual angular difference between nodes $s$ and $t$ falls below this threshold, they have the capability to form connections based on the model formulation (see Eq.(\ref{eq:heaviside})). One the other hand, we can define $\varphi^{\text{e}}(q):=2\pi/q$ as well, being the angular 
scope of a single community. Contrarily to $\varphi^{\text{c}}_{st}$, the latter quantity captures the average width of a circular sector within which a given node can attract intra-community links. For a new-coming node $t$, the number of intra-community links are entirely determined by the relationship between the above two characteristic quantities (see Fig.\ref{fig:intra_connection_ranges_illustration} for details). 

If the effective connection range is greater than the angular 
scope of the communities, i.e. $\varphi^{\text{c}}_{st}>\varphi^{\text{e}}(q)$, then the angular sector associated with the community of node $t$ is completely enclosed by the hyperbolic disc of radius $R_t$ centered around this point (see Fig.\ref{fig:intra_connection_ranges_illustration}b). In this case, the necessary and sufficient condition for node $t$ to receive an intra-community link from $s$ is that $s$ and $t$ falls within the same angular sector, which occurs with probability $q^{-1}$ as given by the second case of Eq.(\ref{eq:pi_cases}).
Conversely, when $\varphi^{\text{c}}_{st}<\varphi^{\text{e}}(q)$, the aforementioned criterion falls short. This is because there are always points within the angular sector corresponding to the community of $t$ that lie outside the hyperbolic disk of radius $R_t$ around this node (refer to Fig. \ref{fig:intra_connection_ranges_illustration}a). In such cases, $\Pi_q(s,t)$ is given by the area of overlap between the angular sector and the hyperbolic disk centered at node $t$ (see the first condition in Eq.(\ref{eq:pi_cases})). 

In the following section, we demonstrate evidence that this process of intra-community link formation is a direct consequence of a specific non-linear preferential attachment rule.

\subsubsection{Non-linear preferential attachment}
 
At first glance, both Eq.(\ref{eq:pi_cases}) and the process of intra-community link formation described above may appear completely unrelated to preferential attachment (PA)~\cite{Laci_science}. Nevertheless, as demonstrated in \ref{subsec:NLPAderiv}, $\Pi_q(s,t)$ appearing in  Eq.(\ref{eq:pi_cases})  can be transformed 
into a form that depends on the degree $k_s$ of node $s$ in a non-linear way. To be more specific, Eq.(\ref{eq:pi_cases}) is equivalent to imposing
\begin{align}
\Pi_q(s,t)=\Pi_q(\bar{k}_s,t) \approx \left\{ \begin{array}{lll}
\frac{\bar{k}_s+A_t}{\sum\limits^t_{s=1}\bar{k}_s+A_t}- \frac{q}{4}\left ( \frac{\bar{k}_s+A_t}{\sum\limits^t_{s=1}\bar{k}_s+A_t} \right )^2,\ \ \  \text{if} \ \ \ \frac{2}{X(s,t)}< \frac{2\pi}{q},\\
\\
q^{-1},\ \ \  \text{if} \ \ \ \frac{2}{X(s,t)}\geq \frac{2\pi}{q},
\end{array}\right.
\label{eq:NLPA}
\end{align}
where $\bar{k}_s$ is the expected degree of node $s$ at time step $t$ and $A_t$ is a variable depending on the model parameters and the current time step $t$ (details are given in \ref{subsec:NLPAderiv}). 
The form of Eq.~(\ref{eq:NLPA}) suggests that the establishment of intra-community links follows 
a preferential attachment rule with respect to the node degrees (first term on the rhs. of Eq.(\ref{eq:NLPA}) augmented with a quadratic term (second term on the rhs. of Eq.(\ref{eq:NLPA})).

Such non-linearity results in  
an intricate connection between the expected number of intra-community links connected to $s$, denoted by 
$\bar{b}_s(t)$, and both the number of communities, $q$, and the node birth time, $s$. As a result, this dependence also 
affects the results we gain about the expected modularity, $\bar{Q}$. A comprehensive analysis of this intricate relationship is postponed to the forthcoming sections. Meanwhile, the detailed derivation of Eq.(\ref{eq:NLPA}) along with the definition of $A_t$ has been relocated to \ref{subsec:NLPAderiv}.


\subsection{Analytical results on the modularity of ultracold PSO networks}

Comprehensive analytical calculations (shown in \ref{sec:appendixB}) reveal that the expected modularity of PSO networks in the ultracold limit, i.e. when $T$ approaches 0, can be expressed at any time step $t$ as follows
\begin{equation}
\lim\limits_{T\rightarrow 0}\bar{Q}\approx \sum\limits_{s=1}^{t}\frac{\bar{b}_s(t)}{2mt}-\frac{1}{q}:=\sum\limits_{s=1}^{t}\frac{I_1(t)+I_2(t)}{2mt}-\frac{1}{q},
    \label{eq:QI1I2}
\end{equation}
where 
\begin{equation}
        I_1(t) = \int\limits^{s}_{1}\Pi_q(i,s)\mathrm{d}i
\end{equation}
\begin{equation}
    =\left\{\begin{matrix}
I_{1a}:=m-q\frac{m^2}{4}\left ( \frac{1-\beta}{s^{1-\beta}-1} \right )^2\frac{s^{1-2\beta}-1}{1-2\beta},{\color{black}\ \ \ \text{if}\ \ \ s'<1},\\ 
I_{1b}:=\frac{s'-1}{q}+\frac{m(1-\beta)}{s^{1-\beta}-1}\frac{s^{1-\beta}-(s')^{1-\beta}}{1-\beta}-q\frac{m^2}{4}\left ( \frac{1-\beta}{s^{1-\beta}-1} \right )^2\frac{s^{1-2\beta}-(s')^{1-2\beta}}{1-2\beta},{\color{black}\ \ \ \text{if}\ \ \ 1<s'<s},
\\ 
I_{1c}:=\frac{s-1}{q}, {\color{black} \ \ \ \text{if}\ \ \ s'>s},
\end{matrix}\right.
\label{eq:I1_global}
\end{equation}
with $s'=\left ( \frac{qm}{2} \frac{1-\beta}{s^{1-\beta}-1}\right )^{1/\beta}$, and
\begin{equation}
I_2(t)=\int\limits^{t}_{s}\Pi_q(i,s)\mathrm{d}i
\end{equation}
\begin{equation}
        = \left\{\begin{matrix}
I_{2a}:=\frac{m}{I_t}s^{-\beta}\frac{t^\beta-s^{\beta}}{\beta}-q\frac{m^2}{4I^2_t}s^{-2\beta}\frac{t^{2\beta-1}-s^{2\beta-1}}{2\beta-1}{\color{black}\ \ \ \text{if}\ \ \ s''<s},\\ 
I_{2b}:=\frac{s''-s}{q}+\frac{m}{I_t}s^{-\beta}\frac{t^\beta-(s'')^{\beta}}{\beta}-q\frac{m^2}{4I^2_t}s^{-2\beta}\frac{t^{2\beta-1}-(s'')^{2\beta-1}}{2\beta-1}{\color{black}\ \ \ \text{if}\ \ \ s<s''<t},
\\ 
I_{2c}:=\frac{t-s}{q} {\color{black} \ \ \ \text{if}\ \ \ s''>t},
\end{matrix}\right.
\label{eq:I2_global}
\end{equation}
where $s''=\left ( 1+ \frac{qm}{2} \frac{1-\beta}{s^{\beta}}\right )^{1/(1-\beta)}$ and $I_t=\frac{1-t^{-(1-\beta)}}{1-\beta}$. 
For the sake of notational simplicity, we omitted to explicitly indicate the dependence of $\bar{Q}, I_1, I_2$ on the additional parameters $\beta,m,q$ in Eqs.(\ref{eq:QI1I2}-\ref{eq:I2_global}). 
Besides, note that the expressions in Eq.(\ref{eq:I1_global}) and Eq.(\ref{eq:I2_global}) are applicable for all values of the popularity fading, denoted as $\beta$, excluding the set $\{0,\frac{1}{2},1\}$. To maintain simplicity, the formulas for these degenerate cases where $\beta$ belongs to the set $\{0,\frac{1}{2},1\}$ are handled separately, and moved to \ref{subsec:expbs} along with the detailed derivations of Eq.(\ref{eq:I1_global}) and Eq.(\ref{eq:I2_global}). 

\begin{figure}[hbt!]
\includegraphics[width=1.\textwidth]{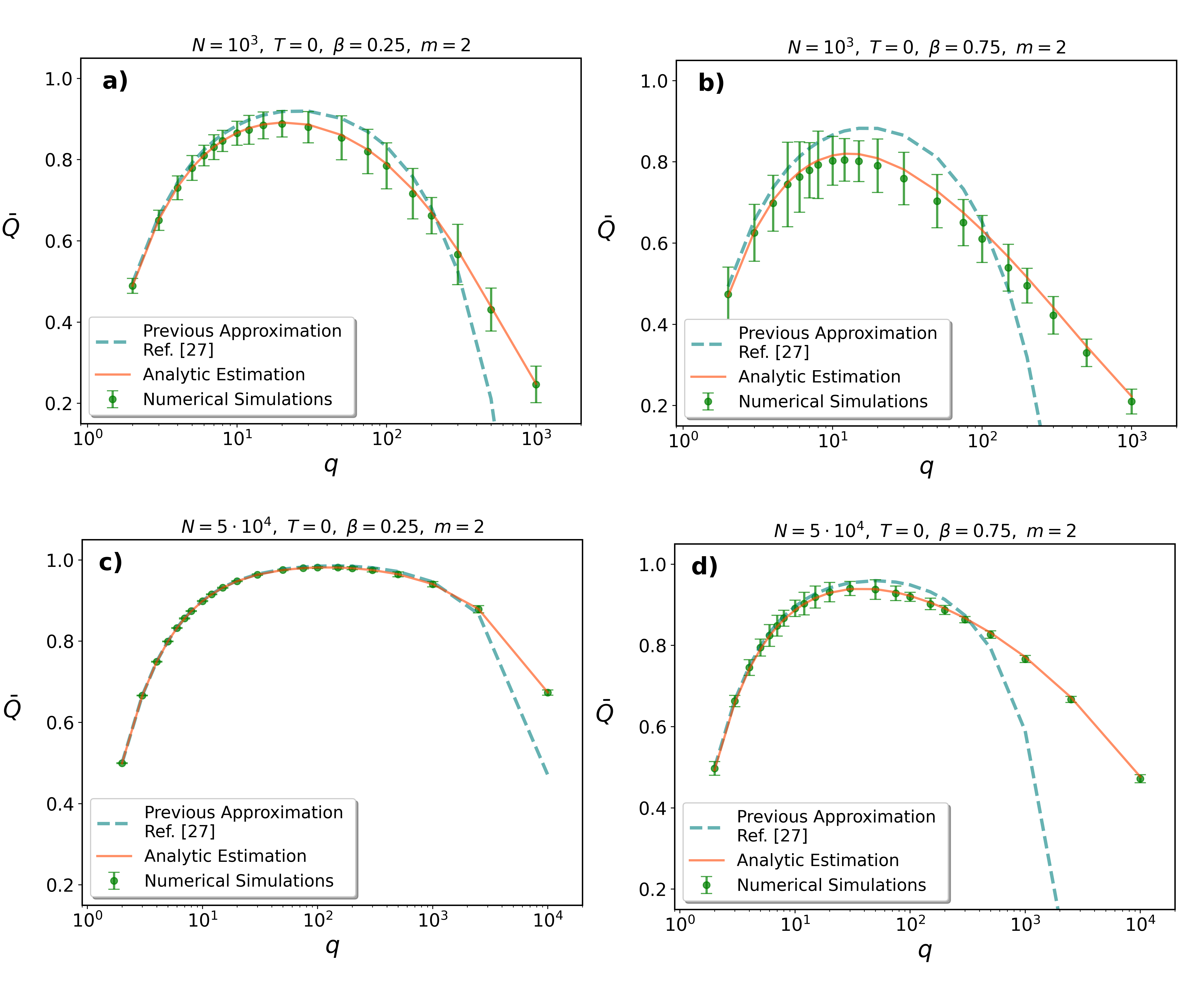}
\caption{\textbf{Expected modularity $\bar{Q}$ in ultracold ($T=0$) PSO networks as a function of the number of modules $q$, for different values of the network size $N$ and the popularity fading paramater $\beta$. } In panels \textbf{a)}-\textbf{d)}, theoretical modularity curves are depicted in orange, accompanied by the exact modularity values averaged over the studied PSO networks shown as green circles. 
Error bars denoting $\pm 5\sigma$ standard deviations are also included for reference.
Additionally, analytical predictions from Ref.~\cite{Balogh2023} are highlighted by dashed lines in a teal color. The number of samples considered in each panel \textbf{a)}-\textbf{d)} is set to 25.
\label{fig:validation_simulation}}
\end{figure}

Before delving into a more in-depth analysis of the aforementioned results, let us examine how well Eq.(\ref{eq:QI1I2}) approximates the average modularity measured in networks generated by the ultracold PSO model with uniform angular partitioning into $q$ communities. In Fig. \ref{fig:validation_simulation}a)-d), we illustrate the dependence on $q$ for both the measured and expected modularity, computed at various model parameters according to Eq.(\ref{eq:Q_exact}) and Eq.(\ref{eq:QI1I2}), respectively. Furthermore, to ensure comparability, a less accurate reference estimation is also included in each panel a)-d), as developped in the study by Balogh \textit{et al.}~\cite{Balogh2023}. Collectively, these results clearly indicate that although the previous estimation of Ref.~\cite{Balogh2023} works reasonably well, especially in the $q\ll N$ regime, Eq.(\ref{eq:QI1I2}) offers a significantly better approximation, as evidenced by the complete overlaps between the numerical and analytical curves in Fig.\ref{fig:validation_simulation}c)-d). Another noteworthy characteristic of our prediction in Eq.(\ref{eq:QI1I2}) is its consistent and maintained accuracy, both for small networks and when dealing with large values of $q$.

\begin{figure}[hbt]
\includegraphics[width=1.\textwidth]{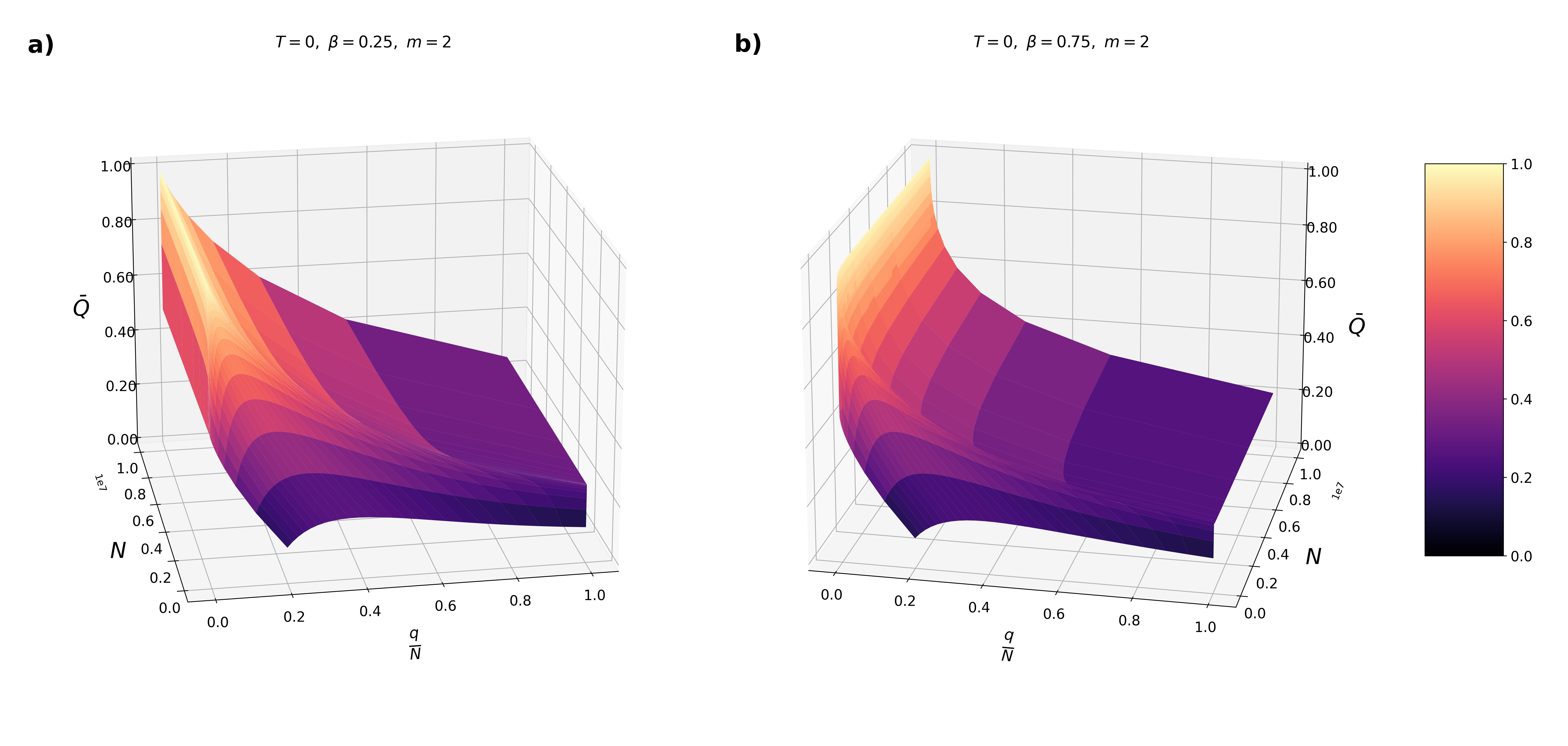}
\caption{\textbf{
Three-dimensional surfaces of the expected modularity in PSO networks as a function of $N$ and $q/N$ for different values of the $\beta$ parameter.
} In panels \textbf{a)}-\textbf{d)}, we depict the predictions of Eq.(\ref{eq:QI1I2}) with the help of a colorbar on the right, showcasing that $\bar{Q}$ approaches its maximum value of $1$ for large networks. 
\label{fig:mod_surface}}
\end{figure}

As a supplement to the previous validation results, Fig.\ref{fig:mod_surface}a)-b) include three-dimensional modularity surfaces over the $N$ and $q/N$ variables for different $\beta$ values. These visualization vividly illustrate the rapid convergence of modularity towards its maximum value of 1, particularly notable for small $\beta$ values. This significantly refined accuracy measurement reaffirms and extends the results of Ref.~\cite{Balogh2023} for $T=0$.

While deriving a closed-form expression for the modularity in Eq.(\ref{eq:QI1I2}) is generally impossible, a relatively simple formula can still be obtained in the limit of $T =\beta = 0$. In the following sections, we examine this limiting scenario and demonstrate that PSO networks degenerate to Random Geometric Graphs~\cite{penrose_rgg} under these specific conditions. Additionally, we conduct a comprehensive investigation into the circumstances that lead to maximal modularity in RGGs.

\subsection{The special case of the PSO model at $T=\beta=0$: Random Geometric Graphs }


When alongside the temperature $T$, the popularity fading parameter $\beta$ is also set to $0$, the nodes of the network drift away from the origin with a maximum possible speed. In this specific limit, nodes are located at the periphery of the native disk, each maintaining a constant radial coordinate $r_N=R=\frac{2}{\zeta}\ln N$ at the end of the network generation process. Nevertheless, the angular coordinates are still distributed uniformly, and assuming $T=0$, connections between pairs of nodes exist only if their corresponding hyperbolic distance is less than a specified threshold (refer to Eq.(\ref{eq:heaviside}) for more details). Since the radial coordinates become identical for all nodes in this limit, this threshold in the hyperbolic distance can be converted into a threshold in the angular difference and hence, the generated network becomes equivalent to a random geometric graph defined on the unit circle (where nodes are distributed uniformly at random angles, and become connected if the angular distance is below the threshold). In order to simplify the forthcoming calculations, we apply a consistent mapping from the unit circle to the unit interval, allowing the condition for connection to be reformulated as 
\begin{equation}
    \frac{1}{2}-\left|\frac{1}{2}-\left|x_s-x_t\right|\right| \leq \frac{\left<k\right>}{4N}
    \label{eq:t0b0_reparametrised}
\end{equation}
where $\left<k\right>$ is the average degree and $s,t$ are arbitrary nodes with their positions in the new coordinate system denoted by $x_s$ and $x_t$, respectively. 
As Eq.(\ref{eq:t0b0_reparametrised}) features a one-dimensional Euclidean $L^1$-norm, the PSO model in this specific setup degenerates to the one-dimensional Random Geometric Graph (RGG) model that is equipped with periodic boundary conditions. A more detailed description of this limit is relocated to \ref{subsec:correspondence}. 


Although it has been previously shown in Ref.~\cite{Balogh2023} that PSO networks can exhibit exceptionally high modularity values, 
the exact conditions leading to this phenomenon remain unidentified.
Motivated by this incompleteness, our upcoming section offers a comprehensive analysis on this matter, specifically focusing on the RGG ($T=\beta=0$) case. Therein, we discuss the roles of both the parameter $q$ and the average degree $\left<k\right>$, as well as their complex interplay through simple scaling arguments.

\subsubsection{Modularity of RGGs}

\begin{figure}[hbt]
\includegraphics[width=1.\textwidth]{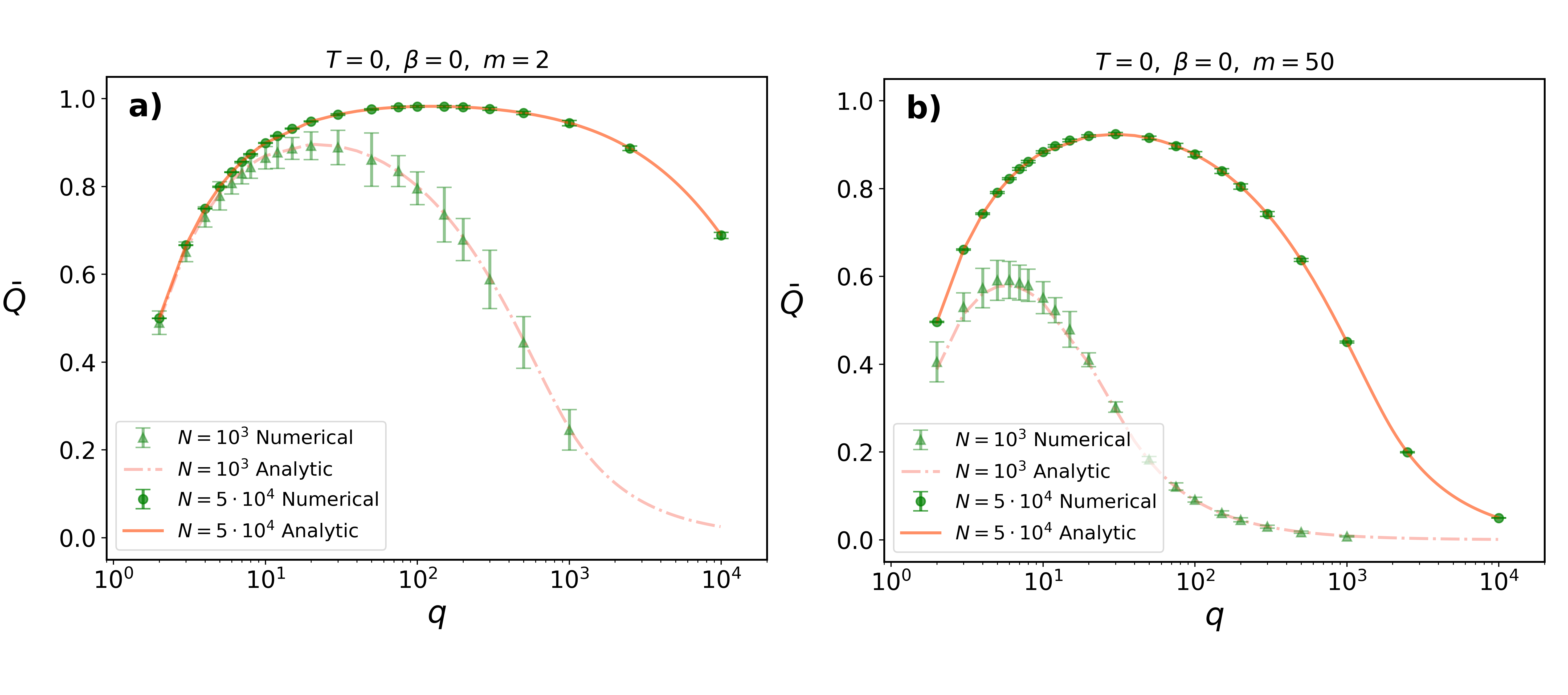}
\caption{\textbf{Expected modularity $\bar{Q}$ in the Random Geometric Graph limit of PSO networks ($T=\beta = 0)$ as a function of $q$ combinations of the parameters $N$ and $m$.}  In panels \textbf{a)}-\textbf{b)}, the solid orange curves represent the theoretical predictions for the modularity values. These are accompanied by exact modularity values, averaged over the studied RGG networks, displayed as green circles. Error bars denoting $\pm 5\sigma$ standard deviations are also included for reference. 
The panels \textbf{a)} and \textbf{b)}, representing $m=2$ and $m=50$, respectively, illustrate the outcomes for networks with dimensions $N=10^3$ and $N=5\cdot 10^4$. The number of samples considered for each curve in both panels \textbf{a)} and \textbf{b)} is consistently set at 25.
\label{fig:RGG_validation_simulation}}
\end{figure}

When the PSO model simplifies to the RGG model, a relatively simple closed-form expression can be derived for the expected modularity as presented in Eq.(\ref{eq:QI1I2}). More specifically, the analytical calculations reveal that $\Bar{Q}$ in this case can be expressed as
\begin{equation}
    \lim\limits_{T,\beta\to 0}\bar{Q}=\bar{Q}_{\text{RGG}}=\left\{\begin{matrix}
1-\frac{qm}{4N} \left ( \ln \frac{N}{1+\frac{qm}{2}} +\frac{3}{2} \right ) -\frac{1}{q}, \ \ \ \text{if} \ \ \ 1+\frac{qm}{2}<N,
\\ 
\frac{N}{2mq}-\frac{1}{q}, \ \ \ \text{if} \ \ \ 1+\frac{qm}{2}\geq N,
\end{matrix}\right.   
    \label{eq:RGGmod}
\end{equation}
the derivation of which can be found in \ref{sec:appendixC}. Before proceeding with the implications of the formula in Eq.(\ref{eq:RGGmod}), let us initially showcase how accurately it predicts the modularity of RGG graphs under the uniform partitioning scheme. In Fig.\ref{fig:RGG_validation_simulation}a)-b), we depict the dependency on $q$ for both the measured modularity (computed using Eq.(\ref{eq:Q_exact})) and the expected modularity (as defined in Eq.(\ref{eq:RGGmod})) for RGG graphs with varying parameters of $m$. As clearly visible in panels a)-b) of Fig.\ref{fig:RGG_validation_simulation}, the theoretical and numerical curves exhibit perfect alignment in each case, emphasizing the exceptional performance of the analytical formula in Eq.(\ref{eq:RGGmod}), even when applied to relatively small networks and large $m$ values.

As a subsequent step, to explore the role of both the graph structure and the partitioning, as well as their interplay, let us consider general graph limits where we let $q\sim N^{\lambda}$ and $\left<k\right>\sim N^{\kappa}$, featuring scaling exponents $\lambda\in [0,1]$ and $\kappa\in [0,2]$. These scaling forms, when combined with Eq.(\ref{eq:RGGmod}), also enable the identification of conditions under which the modularity of RGG graphs can reach high values. 

Let us initiate the analysis with the case corresponding to $\lambda + \kappa > 1$. In this first limit, the asymptotic modularity of RGG graphs typically tends to be very low, usually characterized by negative values. More specifically, if the network is dense, i.e., $\kappa > 1$, or sparse but the number of partitions $q$ scales rapidly enough, then $\lim\limits_{N\to \infty}\bar{Q}_{\text{RGG}}\sim -\frac{1}{N^{\lambda}}$. This approaches $0^{-}$ if $\lambda\neq 0$ and assumes a negative constant value otherwise. 

In a broader context, if the average degree $\left<k\right>$ is sufficiently large or the communities are small enough, then a significant number of nodes $s$ typically connect, on average, to all other nodes in their community, i.e., they satisfy $\bar{b}_s =\frac{N-1}{q}$, resulting in the limitation that modularity cannot exceed the upper limit of $\bar{Q}_{\text{RGG}}=\frac{1}{4}$. In \ref{subsec:scaling}, we demonstrate that this phenomenon takes place when $\frac{qm}{2}(1-\beta)>N$ for $\beta\in (0,1)$ and $\frac{qm}{2}> N\ln N $ for $\beta = 1$. Further details and the complete derivation of the previous results are also found in \ref{subsec:scaling}.

Contrarily to the preceding cases, when $\lambda + \kappa = 1$ the asymptotic modularity of RGG graphs can already be very high. For $\lambda \in (0,1]$ and $1+\frac{qm}{2}<N$, the minimal asymptotic modularity attainable in this limit is $\bar{Q}^{\text{min}}_{\text{RGG}} =\frac{1}{4}$, while the maximal $\bar{Q}_{\text{RGG}}$ can approach values arbitrarily close to the upper limit of $1$. For further details refer to \ref{subsec:scaling}.

As the value of $\lambda + \kappa$ falls below the critical threshold of $1$ and $\lambda > 0$, we enter a regime where Random Geometric Graphs exhibit exceptionally large modularity values. In this highly modular limit, the networks must consistently remain sparse, signifying that $\kappa < 1$, and the number of communities $q$ should scale slower than $N^{1-\kappa}$. Until these conditions are met, ultracold PSO networks with $\beta=0$ are consistently maximally modular, meaning that $\lim\limits_{N\to \infty}\bar{Q}_{\text{RGG}}=1$. Nevertheless, in cases where $\kappa<1$ and $\lambda=0$, RGG graphs are also characterized by high modularity values, albeit never maximal. In this scenario, the precise limit is given by $\lim\limits_{N\to \infty}\bar{Q}_{\text{RGG}}=1-\frac{1}{q}$, similarly to the result obtained in Ref.~\cite{Balogh2023}.

\section{Discussion and conclusions}

The PSO model\cite{PSO} provides an intuitive and concise approach for modelling growing systems where the generated random graphs can reproduce the small-world, scale-free, highly clustered and also strongly modular structure of real-world complex networks in a natural way. The nature of the modular structure of PSO networks has been in the focus of several recent works focusing on hyperbolic networks. 
One of the surprising results was that when partitioned into equal angular sectors, the modularity of PSO networks can approach 1 in thermodynamic limit.

In close relation with the above advancements, in the present work we developed an analytical framework specifically tailored for ultracold PSO networks at $T=0$, which enables remarkably accurate modularity estimates. In particular, we adopted a 
simple yet very informative partitioning scheme by dividing the entire hyperbolic disk into $q$ equally-sized circular sectors, defining the communities accordingly. This approach is similar to the ones employed in Refs.~\cite{commSector_commDetMethod, Balogh2023, Mod_of_RHG}.
While the modularity values obtained in this approach may usually remain below those achieved by state-of-the-art community finding methods, they consistently converge, in expectation, to the maximum possible value of 1, even for medium-sized networks. Remarkably, this provides another strict evidence that hyperbolic networks reach a highly modular state in the thermodynamic limit, which is characterized by maximal modularity~\cite{Balogh2023,Mod_of_RHG}.

Another notable advancement of this article compared to Ref.~\cite{Balogh2023} is that the formulas in Eqs.(\ref{eq:QI1I2}-\ref{eq:I2_global}) allow for a much more accurate estimation of $\bar{Q}$, applicable for any arbitrary values of $q$ and $m$. As a result, modularity can reliably estimated even if, for instance, $q$ is very large approaching the resolution limit $q_{\text{res}}$, or the network is extremely dense $\left<k\right> \sim N^{\lambda}$ with $\lambda>1$, all of which are challenging cases, where the formulas in Ref.~\cite{Balogh2023} fail to predict reliable results.

Regrettably, providing closed-form expressions for the analytical formulas presented in this paper are probably not feasible in general, except for the special case where, alongside the temperature $T$, the parameter $\beta$ governing the popularity fading of the nodes is also set to zero. In this specific limit the PSO model degenerates to the well-know Random Geometric Graph model~\cite{penrose_rgg}, where the expected modularity is given by
\begin{equation}
    \bar{Q}_{\text{RGG}}=\left\{\begin{matrix}
1-\frac{q\left<k\right>}{8N} \left ( \ln \frac{N}{1+\frac{q\left<k\right>}{4}} +\frac{3}{2} \right ) -\frac{1}{q}, \ \ \ \text{if} \ \ \ 1+\frac{q\left<k\right>}{4}<N,
\\ 
\frac{N}{q\left<k\right>}-\frac{1}{q}, \ \ \ \text{if} \ \ \ 1+\frac{q\left<k\right>}{4}\geq N,
\end{matrix}\right.   
    \label{eq:RGGmod_final}
\end{equation}
assuming the average degree of the network is $\left<k\right>$, and the network is partitioned into $q$ number of communities. In the limit of small number of communities, applying the first-order Taylor expansion of the formula in Eq.(\ref{eq:RGGmod_final}) around the point $\frac{q}{N}\ll 1$ yields an asymptotic result of
\begin{equation}
    \bar{Q}_{\text{RGG}}\left(\frac{q}{N}\ll 1\right)=1-\frac{\left<k\right>}{8}\ln N\frac{q}{N}-\frac{1}{q} + \mathcal{O}\left(\left(\frac{q}{N}\right)^2\right),
\end{equation}
which, up to constant factors, coincides with the prediction of Ref.~\cite{Balogh2023}. 

As an another important consequence of Eq.(\ref{eq:RGGmod_final}) we observed a qualitatively distinct behavior of $\bar{Q}_{\text{RGG}}$ for dense and sparse RGG networks. In the dense network limit, i.e. when $\left<k(N)\right >\sim N^{\kappa}$ with $\kappa > 1$, maximal modularity can not actually be reached, implying that densification hinders RGG networks to enter a highly modular state. On the other hand, if the networks are sparse the opposite is observed. For $\kappa<1$, RGG graphs are always maximally modular, meaning that if we simultaneously increase both $q(N)$ and the size of the network $N$, we consistently observe the convergence of $\bar{Q}_{\text{RGG}}$ towards 1. Remarkably, this convergence in the thermodynamic limit persists regardless of the rate at which $q(N)$ is increased.

Although the formalism presented here was originally designed for estimating solely the expected modularity in ultracold PSO networks, its applicability extends beyond this specific context. Since the evaluation of modularity relies on computing the expected number of intra-community links, our formalism is equally applicable to any quality score $\delta$ given in the form of
\begin{equation}
    \delta = f\left(\bar{\bm k},\bar{\bm b}, \bm{\varphi}_{q}\right)
\label{eq:gen_quantity}
\end{equation}
where $f$ is an arbitrary function, $\bm{\varphi}_q=[\varphi_1,...,\varphi_q]$ is the vector of central angles corresponding to the communities, while $\bar{\bm k}=[\bar{k}_1,...,\bar{k}_N]$ and $\bar{\bm b}=[\bar{b}_1,...,\bar{b}_N]$ denotes the vector of degrees and intra-community degrees of the nodes, respectively.  A detailed investigation of this observation is beyond the scope of this paper and remains a subject for future research. Nevertheless, the quantities introduced in Eq.(\ref{eq:gen_quantity}), particularly the expected modularity outlined in Eq.(\ref{eq:QI1I2}), which serves as the central point of this study, could potentially enhance our comprehension of community formation within hyperbolic spaces. Additionally, they may aid in the design of efficient hyperbolic embedding algorithms by leveraging the connection between embeddings and community structure~\cite{commSector_hypEmbBasedOnComms_2019, analogyBetweenHypEmbAndComms}.

\section{Acknowledgements}
S. G. B. was supported by the ÚNKP-23-4 New National Excellence Program of the Ministry for Culture and Innovation, from the source of the National Research, Development and Innovation Fund. S. G. B. acknowledges support from the National Laboratory for Health Security, RRF-2.3.1-21-2022-00006. This research was partially supported by the European Union’s Horizon 2020 research and innovation programme under grant agreement no. 101021607, and the National Research, Development and Innovation Office under grant no. K128780. 



\appendix

\section{The PSO model}
\label{sec:appendixA}

\subsection{Native disk representation of the hyperbolic space}
\label{subsec:native}

A commonly employed approach in network science for studying the hyperbolic geometry involves utilizing the native disk representation of the two-dimensional hyperbolic space \cite{hyperGeomBasics}, where the hyperbolic plane of constant curvature $K<0$ is represented by a disk of infinite radius in the Euclidean plane. In contrast to the well-known Poincaré disk model, this representation offers the benefit of having the radial coordinate $r$ of a point (defined as its Euclidean distance from the disk center) match its true hyperbolic distance from the disk center. Moreover, the model is conformal, meaning that Euclidean angles formed by hyperbolic lines exactly correspond to their hyperbolic counterparts.

The hyperbolic distances between the points are always measured along their connecting geodesics, which can either be a hyperbola, or -- if the disk centre falls on an Euclidean straight line connecting the two points -- the corresponding diameter of the disk. Mathematically, the hyperbolic distance $x$ between any two points with polar coordinates $(r,\theta)$ and $(r',\theta')$ can be given by the hyperbolic law of cosines written as
\begin{equation}
    \mathrm{cosh}(\zeta x)=\mathrm{cosh}(\zeta r)\,\mathrm{cosh}(\zeta r')-\mathrm{sinh}(\zeta r)\,\mathrm{sinh}(\zeta r')\,\mathrm{cos}(\Delta\theta),
    \label{eq:hypDist}
\end{equation}
where $\zeta=\sqrt{-K}$ and $\Delta\theta=\pi-|\pi-|\theta-\theta'||$ is the angle between the studied points. Following Ref. \cite{hyperGeomBasics}, for $2\cdot\sqrt{e^{-2\zeta r}+e^{-2\zeta r'}}<\Delta\theta$ and sufficiently large $\zeta r$ and $\zeta r'$, the hyperbolic distance can safely be approximated as
\begin{equation}
    x\approx r+r'+\frac{2}{\zeta}\cdot\ln\left(\frac{\Delta\theta}{2}\right).
    \label{eq:hypDistApprox}
\end{equation}
The native disk representation serves as a fundamental model of the two-dimensional hyperbolic space in network science, since it is adopted in various popular hyperbolic graph models, including e.g., the PSO~\cite{PSO}, nPSO~\cite{nPSO}, and E-PSO~\cite{EPSO_HyperMap} models.

\subsection{Detailed algorithmic description of the PSO model}
\label{subsec:algorithmicPSO}

The Popularity-Similarity Optimisation model is a growing hyperbolic network model, which simulates how networks evolve in time over the native disk representation of the two-dimensional hyperbolic space~\cite{PSO}. The free parameters of the model are the total number of nodes $N$, the expected number of connections $m$ that a newly added node establishes, the popularity fading parameter $\beta\in(0,1]$ controlling the speed of the outward drift of the nodes during the network growth, and the temperature $T\in[0,1)$ parameter regulating the cutoff in the connection probability. According to Ref.~\cite{PSO} the $m$ parameter is related to the average degree of the generated networks as $\bar{k}=2m$, while the popularity fading parameter $\beta$ is controlling the degree distribution via the following scaling relation
\begin{equation}
    \lim_{k,N\to \infty }p(k) \sim k^{-1-\frac{1}{\beta}}.
\end{equation}
The detailed description of the network generation rules along with the role of each parameter are outlined below. 
\begin{enumerate}
    \item In the $t$-th time step, the new node $t$ appears on the native disk with a radial coordinate $r_{t}(t) = r_t = \frac{2}{\zeta}\ln t$, and an angular coordinate $\theta_t$ sampled from $[0,2\pi)$ uniformly at random. 
    The the curvature $K$ of the two-dimensional hyperbolic space is related to the $\zeta$ parameter as $K=-\zeta^2$, the custom value of which is usually set to $\zeta = 1$.
    \item At each time step, the radial coordinates of all previously added nodes ($s<t$) are increased as $r_{s}(t) = \beta r_{s}+(1-\beta)r_{t}$. This radial outward drift of the nodes is often referred to as 'popularity fading', as it moderates the differences in the nodes' attractivity.
    \item The new node $t$ gets connected to already existing nodes according to the following rules: 
    \begin{itemize}
        \item[a)] If $T=0$, then node $t$ connects to all existing nodes $s<t$ whose hyperbolic distance $x_{st}$ from node $t$ is not larger than
        \begin{equation}
            R_t = \left\lbrace \begin{array}{ll} 
            r_{t}-\frac{2}{\zeta}\ln\left(\frac{2}{\pi}\cdot
             \frac{1-\mathrm{e}^{-\frac{\zeta}{2}(1-\beta)r_{t}}}{m(1-\beta)}\right) & \mathrm{if}\;\; \beta < 1, \\
             r_{t} - \frac{2}{\zeta}\ln\left(\frac{\zeta r_{t}}{\pi\cdot m}\right) & \mathrm{if} \;\; \beta=1. \end{array} \right.
             \label{eq:cutoff_T0}
        \end{equation}
        \item[b)] If $T>0$, then node $t$ gets connected to a previously added nodes $s<t$ with a probability depending on their respective hyperbolic distance $x_{st}$ as
        \begin{equation}
            p(x_{st})=\frac{1}{1+\mathrm{e}^{\frac{\zeta}{2T}(x_{st}-R_t)}}, \label{eq:PSO_link_prob}
        \end{equation}
        where the cut-off value $R_t$ for node $t$ can be calculated as
        \begin{equation}
            R_t = \left\lbrace \begin{array}{ll} 
            r_{t}-\frac{2}{\zeta}\ln\left(\frac{2T}{\sin(T\pi)}\cdot
             \frac{1-\mathrm{e}^{-\frac{\zeta}{2}(1-\beta)r_{t}}}{m(1-\beta)}\right) & \mathrm{if}\;\; \beta < 1, \\
             r_{t} - \frac{2}{\zeta}\ln\left(\frac{T}{\sin(T\pi)}\cdot \frac{\zeta r_{t}}{m}\right) & \mathrm{if} \;\; \beta=1. \end{array} \right.
             \label{eq:cutoff}
        \end{equation}
    \end{itemize}
\end{enumerate}

\subsubsection{Connection probability in the PSO model at $T=0$}
\label{subsubsec:PSOT0connection}

Let us set $\zeta = 1$ and rewrite the distance dependent connection probability in Eq.(\ref{eq:PSO_con_prob}) as
\begin{equation}
     p(x_{st})=\frac{1}{1+e^{\frac{\zeta}{2T}(x_{st}-R_t)}} = \frac{1}{1+\left(X(s,t)\frac{\Delta\theta_{st}}{2}\right)^{\frac{1}{T}}}
     \label{eq:T0limit}
\end{equation}
where we introduced the new variable $X(s,t) =\frac{2}{\pi}s^{\beta}\frac{t^{1-\beta}-1}{m(1-\beta)}$ for the sake of simplicity~\cite{PSO}. By taking the $T\to 0$ in Eq.(\ref{eq:T0limit}) we obtain the definition the reversed Heaviside function given by
\begin{equation}
    \lim\limits_{T\to 0}p(x_{st})=\Theta\left(\Delta\theta_{st}-\frac{2}{X(s,t)}\right)=\left\{ \begin{array}{lll}
1,\ \text{if} \ \Delta\theta_{st}\leq\frac{2}{X(s,t)}\\0, \ \text{otherwise,}
\end{array}\right.
\end{equation}
from which we can immediately recover the formula in Eq.(\ref{eq:heaviside}).

\subsection{The correspondence between the RGG and the PSO model at $T=\beta=0$}
\label{subsec:correspondence}

As outlined in the main text, when setting $T=\beta=0$, the nodes are located at the periphery of the native disk each sharing the same radial coordinate $r_N=R=\frac{2}{\zeta}\ln N$ at the end of the network generation process. Substituting these coordinates back into Eq.(\ref{eq:hypDist}), the hyperbolic distance $x_{st}$ between any pair of nodes $s,t$ can be expressed as
\begin{equation}
    x_{st}\approx 2\frac{2}{\zeta}\ln N+\frac{2}{\zeta}\ln \left(\frac{\Delta\theta_{st}}{2}\right)=\frac{2}{\zeta}\ln\left(\frac{N^2\Delta\theta_{st}}{2}\right).
\end{equation}
In this specific case, the connectivity cut-off in Eq.(\ref{eq:cutoff_T0}) simplifies significantly as well, becoming
\begin{equation}
    R_N = \frac{2}{\zeta}\ln N-\frac{2}{\zeta}\ln \left(\frac{2}{\pi}\frac{1-e^{-\frac{\zeta}{2}\frac{2}{\zeta}\ln N}}{m}\right)\approx \frac{2}{\zeta}\ln \left(\frac{\pi mN}{2}\right),
\end{equation}
based on which the condition for connection between nodes $s,t$ 
can be reformulated as
\begin{align}
    x_{st} = \frac{2}{\zeta}\ln\left(\frac{N^2\Delta\theta_{st}}{2}\right)< \frac{2}{\zeta}\ln \left(\frac{\pi mN}{2}\right) = R_N.
    \label{eq:rgg_connection_ref1}
\end{align}
By exploiting the definition of $\Delta\theta_{st}$, the inequality appearing in Eq.(\ref{eq:rgg_connection_ref1}) can be further simplified to
\begin{equation}
    \pi -\left |\pi - \left|\theta_s - \theta_t \right| \right| <\frac{\pi m}{N},
\end{equation}
which after the reparametrization of the angular coordinates as $x_s = \frac{\theta_s}{2\pi}$ for all nodes $s=1,...,N$, yields 
\begin{equation}
    \frac{1}{2} -\left |\frac{1}{2} - \left|x_s - x_t \right| \right| <\frac{m}{2N}\approx \frac{\left<k\right>}{4N}.
\end{equation}
from which we immediately recover Eq.(\ref{eq:t0b0_reparametrised}).

\section{Intra-community link formation in ultracold PSO networks}
\label{sec:appendixB}

\subsection{Intra-community link formation probability}
\label{subsec:linkformationprob}

Let us suppose that the entire hyperbolic plane is divided into $q$ equally sized angular sectors, each with a central angle of $2\pi/q$. The boundary between the first and second community can be positioned at any arbitrary angle $\alpha$, while the remaining community boundaries are situated at $\alpha+\frac{2\pi}{q}s$, where $s=1,...,q-1$. 
It is apparent that the modularity $Q$ is influenced by the selected value of $\alpha$. Nevertheless, the impact of fluctuations stemming from this choice diminishes in the thermodynamic limit as it is shown in Ref.~\cite{Balogh2023}. Consequently, in what follows, we consistently set $\alpha=0$ for the sake of simplicity. 

Since we define communities as distinct angular sectors on the hyperbolic disk, intra-community links correspond to connections between nodes situated within the same sector. Given this, we can represent the probability that node $s$ establishes an intra-community link with a newly arriving node $t$, assuming both nodes are within community $c$, as follows
\begin{equation}
    \Pi_q(s,t|s,t\in c) =\int\limits_0^{2\pi/q} \varrho\left(\Delta\theta_{st}|s,t\in c\right)p(x_{st})  \,\mathrm{d}\Delta\theta_{st},
    \label{eq:conditional_intra}
\end{equation} 
Considering there are a total number of $q$ communities, the overall probability that any pair of nodes $s,t$ shares an intra-community link in any community can be expressed as
\begin{equation}
\Pi_q(s,t)=\sum\limits^{q}_{c=1}\Pi_q(s,t|s,t\in c )P(s,t\in c ),
    \label{eq:total_law_of_probs}
\end{equation} 
where $P(s,t\in c)=1/q^2$  represents the probability that both nodes $s$ and $t$ belong to sector $c$. Combining Eqs.(\ref{eq:heaviside}-\ref{eq:intra_angular_diff}) with Eqs.(\ref{eq:conditional_intra}-\ref{eq:total_law_of_probs}), and exploiting the statistical equivalence of communities, we obtain
\begin{align}
\Pi_q(s,t)&=\sum\limits^{q}_{c=1}\left(\int\limits^{2\pi/q}_{0}\varrho\left(\Delta\theta_{st}|s,t\in c\right)p(x_{st})  \,\mathrm{d}\Delta\theta_{st}\right) P(s,t\in c )
    \label{eq:piq_derivation}
    \\
    & = \sum\limits^{q}_{c=1}\frac{1}{q^2}\int\limits^{2\pi/q}_{0}\left(\frac{q}{\pi}-\frac{1}{2}\left(\frac{q}{\pi}\right)^2\Delta\theta_{st}\right)\Theta\left(\Delta\theta_{st}-\frac{2}{X(s,t)}\right)\mathrm{d}\Delta\theta_{st}
    \\
    & = \frac{1}{q}\int\limits^{2\pi/q}_{0}\left(\frac{q}{\pi}-\frac{1}{2}\left(\frac{q}{\pi}\right)^2\Delta\theta_{st}\right)\Theta\left(\Delta\theta_{st}-\frac{2}{X(s,t)}\right)\mathrm{d}\Delta\theta_{st}
    \\
    & = \left\{ \begin{array}{lll}
    \int\limits^{2/X}_{0}\left(\frac{1}{\pi}-\frac{q}{2\pi^2}\Delta\theta_{st}\right)\mathrm{d}\Delta\theta_{st},\ \ \  \text{if} \ \ \ \frac{2}{X(s,t)}< \frac{2\pi}{q},\\
    \\
    \int\limits^{2\pi/q}_{0}\left(\frac{1}{\pi}-\frac{q}{2\pi^2}\Delta\theta_{st}\right)\mathrm{d}\Delta\theta_{st},\ \ \  \text{if} \ \ \ \frac{2}{X(s,t)}\geq \frac{2\pi}{q},
    \end{array}\right..
    \\
    & = \left\{ \begin{array}{lll}
    \frac{2}{\pi }\frac{1}{X(s,t)}-\frac{q}{4\pi^2}\left ( \frac{2}{X(s,t)} \right )^2,\ \ \  \text{if} \ \ \ \frac{2}{X(s,t)}< \frac{2\pi}{q},\\
    \\
    \frac{2}{q}-\frac{q}{4\pi^2}\left ( \frac{2\pi}{q} \right )^2,\ \ \  \text{if} \ \ \ \frac{2}{X(s,t)}\geq \frac{2\pi}{q},
\end{array}\right.
    \\
    & = \left\{ \begin{array}{lll}
\frac{2}{\pi}\frac{1}{ X(s,t)}-\frac{q}{\pi^2}\frac{1}{X^2(s,t)},\ \ \  \text{if} \ \ \ \frac{2}{X(s,t)}< \frac{2\pi}{q},\\
\\
q^{-1},\ \ \  \text{if} \ \ \ \frac{2}{X(s,t)}\geq \frac{2\pi}{q},
\end{array}\right.
\end{align}
from which we can recover Eq.(\ref{eq:pi_cases}). As a next step, let us unfold the inequality in Eq.(\ref{eq:pi_cases}) as follows
\begin{equation}
        2s^{-\beta} t^{-(1-\beta)} \frac{\pi}{2}\frac{m(1-\beta)}{1-t^{-(1-\beta)}}<\frac{2\pi}{q}
\end{equation}
which can be rearranged as
\begin{equation}
        t'(t)=t':=\left(\frac{qm}{2}\frac{1-\beta}{t^{1-\beta}-1}\right)^{1/\beta}<s.
        \label{eq:t_prime}
\end{equation}
Introducing the variable $t'$ for notational simplicity, this new variable governs the behavior of $\Pi_q(s,t)$ in Eq. (\ref{eq:pi_cases}). A similar variable, controlling the behavior of $\Pi_q(t,s)$, can also be defined, with attention given to the inversion of the arguments of $\Pi_q$. Changing the role of the $s,t$ variables in Eq.(\ref{eq:pi_cases}) we obtain  
\begin{equation}
        2t^{-\beta} s^{-(1-\beta)} \frac{\pi}{2}\frac{m(1-\beta)}{1-s^{-(1-\beta)}}<\frac{2\pi}{q} 
\end{equation}
which upon solving for $s$ once more, yields
\begin{equation}
        t''(t)=t'':=\left(1+\frac{qm}{2}\frac{(1-\beta)}{t^{\beta}}\right)^{1/(1-\beta)}<s,
        \label{eq:t_doubleprime}
\end{equation}
where $t''$ again determines the behavior of the $\Pi_q(t,s)$ function. In the upcoming sections, we will observe that both $t'$ and $t''$ are crucial, as they play an important role in assessing the number of intra-community links, necessitating the evaluation of both $\Pi_q(s,t)$ and its reverse counterpart $\Pi_q(t,s)$.

\subsection{Derivation of the non-linear preferential attachment rule in the intra-community link formation process}
\label{subsec:NLPAderiv}

Let us begin by recalling that the expected degree of node $s$ at time $t$ in PSO networks can be analytically calculated as
\begin{align}
\bar{k}_s(t)&=\int\limits^{s}_{1}\Pi(i,s)\mathrm{d}i+\int\limits^{t}_{s}\Pi(s,j)\mathrm{d}j\approx m+\frac{m}{I_t\beta}\left(\left(\frac{s}{t}\right)^{-\beta}-1\right),
\label{eq:expecteddeginpso}
\end{align}
as demonstrated in~\cite{PSO, EPSO_HyperMap}. After rearranging Eq.(\ref{eq:expecteddeginpso}) we obtain 
\begin{equation}
    \beta(\bar{k}_s(t)-m) +\frac{m}{I_t} = s^{-\beta}t^{\beta}\frac{m}{I_t}.
    \label{eq:NLPAderiv1}
\end{equation}
Next, we proceed by simplifying the following expression
\begin{equation}
    \frac{2}{\pi X(s,t)}=ms^{-\beta}\frac{1-\beta}{t^{1-\beta}-1}\frac{t^{\beta-1}}{t^{\beta-1}} = ms^{-\beta}t^{\beta-1}\frac{1-\beta}{1-t^{-(1-\beta)}} = \frac{1}{t}s^{-\beta}t^{\beta}\frac{m}{I_t},
    \label{eq:NLPAderiv2}
\end{equation}
where we used the definition of $I_t$. Now, comparing Eq.(\ref{eq:NLPAderiv1}) with Eq.(\ref{eq:NLPAderiv2}) and performing simple algebraic manipulations, we derive the following identities
\begin{align}
    \frac{2}{\pi X(s,t)} &= \frac{\beta(\bar{k}_s(t)-m) +\frac{m}{I_t}}{t} = m\frac{\beta(\bar{k}_s(t)-m) +\frac{m}{I_t}}{mt} 
    \\
    &\approx m\frac{\beta(\bar{k}_s(t)-m) +\frac{m}{I_t}}{\sum\limits^t_{s=1}\beta(\bar{k}_s(t)-m) +\frac{m}{I_t}}= m\frac{\bar{k}_s(t)-m +\frac{m}{\beta I_t}}{\sum\limits^t_{s=1}\bar{k}_s(t)-m +\frac{m}{\beta I_t}},
    \label{eq:NLPA_deriv1}
\end{align}
where we exploited that for large values of $t$
\begin{align}
    \sum\limits^t_{s=1}\beta(\bar{k}_s(t)-m) +\frac{m}{I_t}=2\beta E -\beta mt +\frac{mt}{I_t}\approx 2\beta mt -\beta mt +mt(1-\beta)=mt.
\end{align}
By introducing a new variable $A_t$ in Eq.(\ref{eq:NLPA_deriv1}) as
\begin{equation}
    A_t = \frac{m}{\beta I_t} - m
\end{equation}
we recover the non-linear preferential attachement rule in Eq.(\ref{eq:NLPA}).

\subsection{Expected number of intra-community links in ultracold PSO networks}
\label{subsec:expbs}

In PSO networks, the expected number of intra-community links that any node $s$ has by time $t$ can be expressed as
\begin{align}
\bar{b}_s(t)=\int\limits^{s}_{1}\Pi_q(i,s)\mathrm{d}i+\int\limits^{t}_{s}\Pi_q(s,i)\mathrm{d}i :=I_1+I_2,
    \label{eq:expectedintradegpso}
\end{align}
where $\Pi_q(s,t)$ is defined as the overall probability that node $s$ establishes an intra-community link with a newly arriving node $t$, considering the hyperbolic disk divided into $q$ equally sized circular sectors. Note that the dependency of Eq.(\ref{eq:expectedintradegpso}) on both $\Pi_q(i, s)$ and its reverse counterpart $\Pi_q(s, i)$ underscores again the importance of the variables $t'$ and $t''$ given in Eq.(\ref{eq:t_prime}) and Eq.(\ref{eq:t_doubleprime}), respectively.   

The following subsections are dedicated to the detailed assessment of $I_1$ and $I_2$ in Eq.(\ref{eq:QI1I2}) and in Eq.(\ref{eq:expectedintradegpso}) at zero temperature for all potential values of $\beta$ within the range of $[0,1]$, with particular attention given to the degenerate cases when $\beta$ assumes values of $\{0, \frac{1}{2}, 1\}$.

\subsubsection{Evaluation of $I_1$ and $I_2$ for $T=0$ and $\beta \notin \{0, \frac{1}{2}, 1\}$}
\label{subsubsec:expbsb}

Let us begin with the calculation of $I_1$ for $\beta \notin \{0, \frac{1}{2}, 1\}$. Depending on the value of $s'$ in Eq.(\ref{eq:t_prime}), we need to distinguish three distinct cases. In the first case where $s'<1$, based on the definition of $s'$ in Eq.(\ref{eq:t_prime}) the condition $i>s'$ consistently holds. Referring back to Eq.(\ref{eq:pi_cases}), we obtain
\begin{align}
I_1(s'<1)&=I_{1a}=\int\limits^{s}_{1}\Pi_q(i,s)\mathrm{d}i=\int\limits_{1}^{s}\frac{2}{\pi X(i,s)}-\frac{q}{\pi^2X^2(i,s)}\mathrm{d}i\\
&=\frac{m(1-\beta)}{s^{1-\beta}-1}\int\limits_{1}^{s}i^{-\beta}\mathrm{d}i-\frac{m^2q}{4}\left ( \frac{1-\beta}{s^{1-\beta}-1} \right )^2\int\limits_{1}^{s}i^{-2\beta}\mathrm{d}i
\label{eq:I1detailed_first}
\\
&=\frac{m(1-\beta)}{s^{1-\beta}-1}\frac{s^{1-\beta}-1}{1-\beta}-\frac{m^2q}{4}\left ( \frac{1-\beta}{s^{1-\beta}-1} \right )^2\frac{s^{1-2\beta}-1}{1-2\beta}
\\
&=m-\frac{m^2q}{4}\left ( s^{\beta-1}\frac{1-\beta}{1-s^{-(1-\beta)}} \right )^2\frac{s^{1-2\beta}-1}{1-2\beta}
\\
&=m-\frac{m^2q}{4I^2_s}\frac{s^{-1}-s^{2\beta-2}}{1-2\beta}.
\end{align}
The second scenario arises when $1<s'<s$. In contrast to the prior case, Eq.(\ref{eq:pi_cases}) now simplifies into a more intricate form, resulting in
\begin{align}
I_1(1<s'<s)&=I_{1b}=\int\limits^{s}_{1}\Pi_q(i,s)\mathrm{d}i=\int\limits_{1}^{s'}\frac{1}{q}\mathrm{d}i-\int\limits_{s'}^{s}\frac{2}{\pi X(i,s)}-\frac{q}{\pi^2X^2(i,s)}\mathrm{d}i\\
&= \frac{s'-1}{q}+m\frac{s^{1-\beta}-s'^{1-\beta}}{s^{1-\beta}-1}-\frac{qm^2}{4}\left ( \frac{1-\beta}{s^{1-\beta}-1} \right )^2\frac{s^{1-2\beta}-s'^{1-2\beta}}{1-2\beta}.
\end{align}
Lastly, in the scenario where $s'>s$, it is automatically true that $i<s'$. Consequently, one can observe that $\Pi_q(s,t)=1/q$, leading to
\begin{align}
I_1(s<s')&=I_{1c}=\int\limits^{s}_{1}\Pi_q(i,s)\mathrm{d}i=\int\limits_{1}^{s}\frac{1}{q}\mathrm{d}i=\frac{s-1}{q}.  
\end{align}

As a next step, let us proceed with the evaluation of $I_2$. The same approach utilized for the evaluation of $I_1$ is also applicable here, with the only difference being that the value of $I_2$ is determined by $s''$ rather than $s'$ (see Eq.(\ref{eq:t_prime}) and Eq.(\ref{eq:t_doubleprime})). First, if $s''<s$, then $i>s''$, hence $I_2$ reduces to
\begin{align} I_2(s''<s)&=I_{2a}=\int\limits^{t}_{s}\Pi_q(s,i)\mathrm{d}i=\int\limits_{s}^{t}\frac{2}{\pi X(s,i)}-\frac{q}{\pi^2X^2(s,i)}\mathrm{d}i\\
&=  ms^{-\beta}\int\limits^{t}_{s}\frac{i^{-(1-\beta)}}{I_i}\mathrm{d}i-\frac{qm^2s^{-2\beta}}{4}\int\limits^{t}_s \frac{i^{-2(1-\beta)}}{I^2_i}\mathrm{d}i\\
&\approx   \frac{ms^{-\beta}}{I_t}\int\limits^{t}_{s}i^{-(1-\beta)}\mathrm{d}i-\frac{qm^2s^{-2\beta}}{4I^2_t}\int\limits^{t}_s i^{-2(1-\beta)}\mathrm{d}i\\
&=\frac{m}{I_t}\frac{\left ( \frac{s}{t} \right )^{-\beta}-1}{\beta}-\frac{qm}{4I^2_t}s^{-1}\frac{\left ( \frac{s}{t} \right )^{1-2\beta}-1}{2\beta-1},
\end{align}
where we exploited that $I_i \approx I_t$ for sufficiently large values of $i$ and $t$. Similar manipulations apply to the $s<s''<t$ case. More specifically, we obtain
\begin{align}
I_2(s<s''<t)&=I_{2b}=\int\limits^{t}_{s}\Pi_q(s,i)\mathrm{d}i
\\
&=\int\limits^{s''}_{s}\frac{1}{q}\mathrm{d}i+\int\limits_{s''}^t\frac{2}{\pi X(s,i)}-\frac{q}{\pi^2X^2(s,i)}\mathrm{d}i\\
&=\frac{s''-s}{q}+\frac{m}{I_t}s^{-\beta}\frac{t^{\beta}-s''^{\beta}}{\beta}-\frac{qm}{4I^2_t}s^{-2\beta}\frac{t^{1-2\beta}-s''^{1-2\beta}}{2\beta-1}.
\end{align}
Finally, the last case is where $s''>t$, which implies that $\Pi_q(i,s)=1/q$ yielding
\begin{align}
I_1(s''>t)&=I_{2c}=\int\limits^{t}_{s}\Pi_q(i,s)\mathrm{d}i=\int\limits_{s}^{t}\frac{1}{q}\mathrm{d}i=\frac{t-s}{q}.\\ 
\end{align}

\subsubsection{Evaluation of $I_1$ and $I_2$ for $T=\beta=0$ (RGG)}
\label{subsubsec:expbsrgg}

The preceding considerations are not applicable when $\beta = 0$ because $t'$ in Eq.(\ref{eq:t_prime}) is undefined in this particular limit. However, with some additional attention, we can still derive a well-defined formula for $s'$, resulting in
\begin{equation}
    s'= \delta(1+\frac{qm}{2}-s)=\delta(s''-s),
\label{eq:rgg_sprime}
\end{equation}
where $\delta$ denotes the Dirac-delta function, and $s''$ is given by
\begin{equation}
 s''=1+\frac{qm}{2}. 
\label{eq:rgg_sdoubleprime} 
\end{equation}
With the help of these formulas, $I_1$ in this case can be formulated as 
\begin{equation}
    I_1 = \left\{\begin{matrix}
I_{1a}:=m-q\frac{m^2}{4}\frac{1}{s-1},{\color{black}\ \ \ \text{if}\ \ \  s>1+\frac{qm}{2}\ (s'=0)},\\
I_{1b}:=\text{not possible},
\\ 
I_{1c}:=\frac{s-1}{q}, {\color{black}\ \ \ \text{if}\ \ \  1+\frac{qm}{2}>t \ (s'=\infty) \text{ or } s<1+\frac{qm}{2} \ (s'=\infty)}
\end{matrix}\right.
\label{eq:I1_b0}
\end{equation}
or equivalently, in a more compact form as
\begin{equation}
I_1= m-q\frac{m^2}{4}\frac{1}{s-1} \ \ \ \ \forall s.
\label{eq:I1_b0_short}
\end{equation}
In analogous fashion to the derivation of $I_1$, for $I_2$, we acquire
\begin{equation}
        I_2 = \left\{\begin{matrix}
I_{2a}:=\frac{mt}{t-1}\ln \frac{t}{s}-q\frac{(mt)^2}{4(t-1)^2}(s^{-1}-t^{-1}) {\color{black}\ \ \ \text{if}\ \ \  1+\frac{qm}{2}<s},\\ 
I_{2b}:=\frac{1+\frac{qm}{2}-s}{q}+\frac{mt}{t-1}\ln \frac{t}{1+\frac{qm}{2}}-q\frac{(mt)^2}{4(t-1)^2}((1+\frac{qm}{2})^{-1}-t^{-1}){\color{black}\ \ \ \text{if}\ \ \ s< 1+\frac{qm}{2}<t},
\\ 
I_{2c}:=\frac{t-s}{q} {\color{black} \ \ \ \text{if}\ \ \  1+\frac{qm}{2}>t},
\end{matrix}\right.
\label{eq:I2_b0}
\end{equation}

\subsubsection{Evaluation of $I_1$ and $I_2$ for $T=0$ and $\beta=1$}
\label{subsubsec:expbsb1}

Additional computations are needed for the $\beta = 1$ case, as many previously used quantities become divergent under this condition. By considering the limit as $\beta \to 1$, the following formulas can be derived
\begin{equation}
    \lim_{\beta\to 1}I_t =\lim_{\beta\to 1} \frac{1-t^{-(1-\beta)}}{1-\beta}=\lim_{\beta\to 1} \frac{-t^{\beta-1}\ln t}{-1}=\ln t
\end{equation}
\begin{equation}
    \lim_{\beta\to 1}\frac{1}{X(s,t)}=\frac{m\pi}{2}s^{-1}\lim_{\beta\to 1}\frac{1-\beta}{t^{1-\beta}-1}=\frac{m\pi}{2s}\lim_{\beta\to 1}\frac{1}{t^{1-\beta}\ln t}=\frac{m\pi}{2s\ln t},
\end{equation}
\begin{equation}
    \lim_{\beta\to 1}s'=\lim_{\beta\to 1}\left ( \frac{qm}{2} \frac{1-\beta}{s^{1-\beta}-1}\right )^{1/\beta}=\frac{qm}{2\ln s},
\end{equation}
\begin{align}
    \lim_{\beta\to 1}s''&=\lim_{\beta\to 1}\left ( 1+ \frac{qm}{2} \frac{1-\beta}{s^{\beta}}\right )^{1/(1-\beta)}=\lim_{\beta\to 1}\left ( 1+ \frac{qm}{2} \frac{s^{-\beta}}{\frac{1}{1-\beta}}\right )^{1/(1-\beta)}\\
    &=\lim\limits_{\alpha\to\infty}\left ( 1+\frac{qm}{2}\frac{s^{1/\alpha-1}}{\alpha} \right )^\alpha=e^{\frac{qm}{2s}},
\end{align}
where $\alpha = \frac{1}{1-\beta}$ is a constant parameter. Leveraging these formulations yields
\begin{align}
     I_1 = \left\{\begin{matrix}
I_{1a}:=m-q\frac{m^2}{4}\frac{1}{\ln^2 s}\left (1-\frac{1}{s}\right),{\color{black}\ \ \ \text{if}\ \ \  s'<1},\\
I_{1b}:=\frac{\frac{qm}{2\ln s}-1}{q}-m\frac{\ln\left ( s\ \frac{2\ln s}{qm} \right )}{\ln s}-\frac{qm^2}{4}\frac{1}{\ln^2 s}\left ( \frac{2\ln s}{qm}-\frac{1}{s} \right ), {\color{black}\ \ \ \text{if}\ \ \  1<s'<s},
\\ 
I_{1c}:=\frac{s-1}{q}, {\color{black}\ \ \ \text{if}\ \ \  s'>s.}
\end{matrix}\right.
\end{align}
In analogous fashion to the derivation of $I_1$, for $I_2$, we obtain
\begin{align}
     I_2 = \left\{\begin{matrix}
I_{2a}:=\frac{m}{\ln t}(t-s)s^{-1}-\frac{qm^2}{\ln^2 t }s^{-2}(t-s),{\color{black}\ \ \ \text{if}\ \ \  s''<s},\\
I_{2b}:=\frac{e^{\frac{qm}{2s}}-s}{q}+\frac{m}{\ln t}s^{-1}\left ( t- e^{\frac{qm}{2s}}\right )-\frac{qm^2}{4\ln^2 t}s^{-2}\left ( t- e^{\frac{qm}{2s}} \right ), {\color{black}\ \ \ \text{if}\ \ \  s<s''<t},
\\ 
I_{2c}:=\frac{t-s}{q}, {\color{black}\ \ \ \text{if}\ \ \  s''>t.}
\end{matrix}\right.
\end{align}

\subsubsection{Evaluation of $I_1$ and $I_2$ for $T=0$ and $\beta=\frac{1}{2}$}
\label{subsubsec:expbsb05}

For a thorough examination of the $\beta = 1/2$ case, it is essential to determine the following limits
\begin{equation}
    \lim_{\beta\to 1/2} \frac{s^{-1}-s^{2\beta-2}}{1-2\beta}=\lim_{\beta\to 1/2} \frac{-s^{2\beta-2}2\ln s}{-2}=s^{-1}\ln s,
\end{equation}
\begin{equation}
    \lim_{\beta\to 1/2} \frac{s^{1-2\beta}-s'^{1-2\beta}}{1-2\beta}=\lim_{\beta\to 1/2} \frac{s^{1-2\beta}\ln s (-2)-s'^{1-2\beta}\ln s' (-2)}{-2}=\ln s-\ln s' = \ln \frac{s}{s'},
\end{equation}
\begin{equation}
    \lim_{\beta\to 1/2} \frac{t^{1-2\beta}-s''^{1-2\beta}}{1-2\beta}= \ln \frac{s''}{t},
\end{equation}
where $s'=\left(\frac{qm}{2}\frac{1}{\sqrt{s}-1}\right)^2$ and $s''=\left(1+\frac{qm}{4}\frac{1}{\sqrt{s}}\right)^2$. Using the above formulas, we obtain the following results for $I_1$ and $I_2$
\begin{equation}
    I_1(s'<1) = I_{1a} = m-\frac{qm^2}{4}\left(\frac{1/2}{\sqrt{s}-1}\right)^2\ln s
\end{equation}
\begin{equation}
    I_1(1<s'<s) = I_{1b} = \frac{s'-1}{q}+m\frac{s^{1/2}-s'^{1/2}}{s^{1/2}-1}-\frac{qm^2}{4}\left(\frac{1/2}{\sqrt{s}-1}\right)^2\ln \frac{s}{s'}
\end{equation}
\begin{equation}
    I_1(s'>s) =I_{1c} =\frac{s-1}{q}.
\end{equation}
\begin{align}
    I_2(s''<s)&I_{2a} =\approx \frac{ms^{-1/2}}{I_t}\int\limits_{s}^{t}i^{-1/2}\mathrm{d}i-\frac{qm^2}{4I^2_t}s^{-1}\int\limits_{s}^{t}i^{-1}\mathrm{d}i
=\frac{m}{I_t}\frac{\left ( \frac{s}{t} \right )^{-1/2}}{1/2}-\frac{qm^2}{4I^2_t}s^{-1}\ln \frac{t}{s}
\end{align}
\begin{align}
    I_2(s''<s)= I_{2b} =\frac{s''-s}{q}+\frac{m}{I_t}\frac{\left ( \frac{s}{t} \right )^{-1/2}-\left ( \frac{s}{s''} \right )^{-1/2}}{1/2}-\frac{qm^2}{4I^2_t}s^{-1}\ln \frac{s''}{t}
\end{align}
\begin{align}
    I_2(s''>t)= I_{2c} =\frac{t-s}{q}.
    \label{eq:I1detailed_last}
\end{align}

\subsection{Expected modularity in ultracold PSO networks as a function of $I_1$ and $I_2$}

Based on Eqs. (\ref{eq:I1detailed_first})-(\ref{eq:I1detailed_last}), the expected modularity in ultracold PSO networks under the uniform partitioning scheme with $q$ communities can be expressed as:
\begin{equation}
\lim\limits_{T\to 0}\bar{Q}\approx \sum\limits_{s=1}^{t}\frac{\bar{b}_s(t)}{2mt}-\frac{1}{q}=\sum\limits_{s=1}^{t}\frac{I_1(t)+I_2(t)}{2mt}-\frac{1}{q}.
\end{equation}
For the sake of clarity, hereinafter, we consistently assess the modularity at the end of the network generation process, i.e. when $t=N$.

\section{Expected modularity in the RGG limit of PSO networks ($T=\beta = 0)$}
\label{sec:appendixC}

\subsection{Closed-form expression for the modularity of RGGs}
\label{subsubsec:closedformrgg}

As discussed in the main text of the article and shown in Eqs.(\ref{eq:rgg_sprime}-\ref{eq:rgg_sdoubleprime}), as well as in Eqs.(\ref{eq:I1_b0}-\ref{eq:I2_b0}), the expected modularity $\lim\limits_{T,\beta\to 0}=\bar{Q}_{\text{RGG}}$ of random geometric graphs is greatly influenced by the quantity $1+\frac{qm}{2}$. For the sake of simplifying notation, let us introduce a new parameter $r=1+\frac{qm}{2}$ and investigate various scenarios based on the relationship between $r$ and $N$ (refer to Eqs.(\ref{eq:rgg_sprime}-\ref{eq:I2_b0})).
Initially, let us concentrate on the case where $r>N$. Clearly, if $r>N$, Eq.(\ref{eq:I1_b0}) and Eq.(\ref{eq:I2_b0}) imply that $I_1=I_{1c}=\frac{s-1}{q}$ and $I_2=I_{2c}=\frac{N-s}{q}$ for all nodes $s$.
As a result, the modularity $\bar{Q}_{\text{RGG}}$ in this scenario reduces to the following simple form:
\begin{align}
    \bar{Q}_{\text{RGG}}&=\frac{1}{2E}\left(\sum\limits^{N}_{s=1}\frac{s-1}{q}+\frac{N-s}{q}\right) -\frac{1}{q}\approx\frac{N-1}{2mq}-\frac{1}{q}
    \\
    &\approx \frac{N}{4(r-1)}-\frac{1}{q}
    =\frac{N}{2qm}-\frac{1}{q},
    \label{eq:RGGmod_copy0}
\end{align}
where we used the fact that $E\approx mN$. Equation (\ref{eq:RGGmod_copy0}) appears as a first case in Eq.(\ref{eq:RGGmod}).

In order to assess the remaining scenarios, let us begin by rephrasing the positive term of the modularity as
\begin{equation}
    \bar{Q}_{\text{RGG}+} = \sum^{N}_{s=1}\frac{\bar{b}_s}{2E} = \sum^{r}_{s=1}\frac{\bar{b}_s}{2E}+\sum^{N}_{s=r+1}\frac{\bar{b}_s}{2E}:=\bar{Q}^{(1)}_{+}+\bar{Q}^{(2)}_{+},
    \label{eq:Q+12}
\end{equation}
where we split the total sum into two parts; the first term in Eq.(\ref{eq:Q+12}) encompasses nodes, where $s<1+\frac{qm}{2}=r$; while the second term in Eq.(\ref{eq:Q+12}) is running over nodes for which $s\geq r$. First, let us focus on the first term, which by invoking Eq.(\ref{eq:I1_b0}) and Eq.(\ref{eq:I2_b0}) yields
\begin{align}
    \bar{Q}^{(1)}_{+}&=\sum\limits^{r}_{s=1}\frac{s-1}{2Eq}+\frac{r-s}{2Eq}+\frac{mN}{2E(N-1)}\ln \frac{N}{r}-\frac{qm^2N^2}{8E(N-1)^2}\left ( \frac{1}{r}- \frac{1}{N}\right )
    \label{eq:Qplus1_0}
    \\
    &=\left(\frac{r-1}{2Eq}+\frac{mN}{2E(N-1)}\ln \frac{N}{r}-\frac{qm^2N^2}{8E(N-1)^2}\left ( \frac{1}{r}- \frac{1}{N}\right )\right)\left(\sum\limits^{r}_{s=1}1\right)
    \\
    & \approx \frac{r(r-1)}{2mNq}+\frac{mNr}{2mN(N-1)}\ln \frac{N}{r}-\frac{qm^2N^2(r-1)}{8mN(N-1)^2}\left ( \frac{1}{r}- \frac{1}{N}\right ),
    \label{eq:Qplus1_1}
\end{align}
where  we again used that $E\approx mN$. In the last equality of Eqs.(\ref{eq:Qplus1_0}-\ref{eq:Qplus1_1}) we also utilized the fact that, in this particular case, $I_1$ equals $I_{1c}$ and $I_2$ equals $I_{2b}$. Asymptotically, i.e. in the $N\to \infty$ limit, Eq.(\ref{eq:Qplus1_1}) can safely be approximated by
\begin{align}
    \bar{Q}^{(1)}_{+}\approx \frac{r(r-1)}{2mNq}+\frac{r}{2N}\ln \frac{N}{r}-\frac{qmr}{8N}\left ( \frac{1}{r}- \frac{1}{N}\right ),
\end{align}
which after the substitution of $r-1=qm/2$ yields
\begin{align}
    \bar{Q}^{(1)}_{+}\approx \frac{r}{4N}+\frac{r}{2N}\ln \frac{N}{r}-\frac{r(r-1)}{4N}\left ( \frac{1}{r}- \frac{1}{N}\right ).
\end{align}

Now let us go back to Eq.(\ref{eq:Q+12}) and procceed by evaluating $\bar{Q}^{(2)}_{+}$, which denotes the contribution of nodes $s$ for which $s\geq r$. In this case $I_1=I_{1a}$ in Eq.(\ref{eq:I1_b0}), and $I_2=I_{2a}$ in Eq.(\ref{eq:I2_b0}). Considering this, we derive the following:
\begin{align}
    \bar{Q}^{(2)}_{+}&=\frac{1}{2E}\sum_{s=r+1}^{N}m-\frac{qm^2}{4}\frac{1}{s-1}+\frac{mN}{N-1}\ln \frac{N}{s}-\frac{q}{4}\left ( \frac{mN}{N-1} \right )^2\left ( \frac{1}{s}-\frac{1}{N} \right )
    \\
    &\approx \frac{(N-r)}{2N}-\frac{qm}{8N}\left(H_{N-1}-H_{r-1}\right)+\frac{1}{2(N-1)}\left((N-r)\ln N -\ln \prod\limits^{N}_{s=r+1}s\right)
    \nonumber
    \\
    &  \phantom{aaaaaaaaaaaaa} -\frac{q}{8mN}\left ( \frac{mN}{N-1} \right )^2(H_N-H_r-1+\frac{r}{N})
    \\
    &\approx \frac{1}{2}\left(1-\frac{r}{N}\right)-\frac{r-1}{4N}\left(H_{N-1}-H_{r-1}\right)+\frac{1}{2}\left(1-\frac{r}{N}\right)\ln N -\frac{1}{2N}\ln \frac{N!}{r!}
    \nonumber
    \\
    &  \phantom{aaaaaaaaaaaaa}
    -\frac{r-1}{4N}(H_N-H_r-1+\frac{r}{N})
    \\
    &\approx \frac{1}{2}\left(1-\frac{r}{N}\right)(1+\ln N)-\frac{r-1}{4N}\left(H_{N}-H_{r-1}\right) -\frac{1}{2N}\ln \frac{N!}{r!}
    \nonumber
    \\
    &  \phantom{aaaaaaaaaaaaa}
    -\frac{r-1}{4N}(H_N-H_r-1+\frac{r}{N})
\end{align}
where $H_k$ denotes the $k$-th Harmonic number. Bringing all parts together, we arrive at:
\begin{align}
\bar{Q}_{\text{RGG}}&=\bar{Q}_{\text{RGG+}}-\bar{Q}_{\text{RGG-}} \approx \bar{Q}^{(1)}_{+}+\bar{Q}^{(2)}_{+}
-\frac{1}{q} = 
\\
&=\frac{r}{4N}+\frac{r}{2N}\ln \frac{N}{r}-\frac{(r-1)r}{4N}\left ( \frac{1}{r}- \frac{1}{N}\right ) +\frac{1}{2}\left(1-\frac{r}{N}\right)(1+\ln N)
\nonumber
\\
&\phantom{aaaaa} -\frac{r-1}{4N}\left(H_N-H_{r-1}\right) -\frac{1}{2N}\ln \frac{N!}{r!}-\frac{r-1}{4N}(H_N-H_r-1+\frac{r}{N})-\frac{1}{q},
\label{eq:Q++Q-}
\end{align}
where we used that $\bar{Q}_{\text{RGG}}=\bar{Q}_{\text{RGG+}}-\bar{Q}_{\text{RGG-}}$, along with the approximation $\bar{Q}_{\text{RGG-}}\approx\frac{1}{q}$ \cite{Balogh2023}. Collecting similar terms and simplifying Eq.(\ref{eq:Q++Q-}) yields
\begin{align}
    \bar{Q}_{\text{RGG}}&=\frac{1}{2}\left(1-\frac{r}{N}\right)(1+\ln N)-\frac{r-1}{4N}\left(2H_N-H_{r-1}-H_r-1+\frac{r}{N}\right)
    \nonumber
    \\
    & \phantom{aaaaaaaaaaaaa} +\frac{r}{4N}+\frac{r}{2N}\ln \frac{N}{r}
    -\frac{1}{2N}\ln \frac{N!}{r!}-\frac{r-1}{4N}\left(1-\frac{r}{N}\right)-\frac{1}{q}
    \\
    &=\frac{1}{2}\left(1-\frac{r}{N}\right)(1+\ln N)-\frac{r-1}{4N}\left(2H_N-H_{r-1}-H_r\right)
    \nonumber
    \\
    & \phantom{aaaaaaaaaaaaa}
    +\frac{r}{4N}+\frac{r}{2N}\ln \frac{N}{r}-\frac{1}{2N}\ln \frac{N!}{r!}-\frac{1}{q},
\end{align}
which in case of $r\gg 1$ reduces to
\begin{align}
    \bar{Q}_{\text{RGG}}&\approx \frac{1}{2}\left(1-\frac{r}{N}\right)(1+\ln N)-\frac{r-1}{4N}\left(2H_N-H_{r-1}-H_r\right)
    \nonumber
    \\
    & \phantom{aaaaaaaaaaaaa}
    +\frac{r}{4N}+\frac{r}{2N}\ln \frac{N}{r}-\frac{1}{2N}\ln \frac{N!}{r!}-\frac{1}{q}
    \\
    &= \frac{1}{2}\left(1-\frac{r}{N}\right)(1+\ln N)-\frac{r-1}{4N}\left(2H_N-H_{r-1}-H_r-2\ln \frac{N}{r}\right)
    \nonumber
    \\
    & \phantom{aaaaaaaaaaaaa}
    +\frac{r}{4N}+\frac{1}{2N}\ln \frac{N}{r}-\frac{1}{2N}\ln \frac{N!}{r!}-\frac{1}{q}
    \\
    &\approx \frac{1}{2}\left(1-\frac{r}{N}\right)(1+\ln N)+\frac{r}{4N}+\frac{1}{2N}\ln \frac{N}{r}-\frac{1}{2N}\ln \frac{N!}{r!}-\frac{1}{q}
    \label{eq:rgg_almost_final}
\end{align}
where we used the fact that $\lim\limits_{k\to\infty}H_k-\ln k \approx \gamma_{EM} \approx 0.5772156649$ with $\gamma_{EM}$ being the Euler-Mascheroni constant. Now, let's continue by applying the Stirling formula to the factorial in Eq. (\ref{eq:rgg_almost_final}), which, for large values of $r$, results in 
\begin{align}
    \bar{Q}_{\text{RGG}}\approx &\frac{1}{2}\left(1-\frac{r}{N}\right)(1+\ln N)+\frac{r}{4N}+\frac{\ln \frac{N}{r}}{2N}+\frac{\left ( r\ln r-r-N\ln N + N  \right )}{2N} -\frac{1}{q}
\\
&=
\frac{1}{2}\left(1-\frac{r}{N}\right)(1+\ln N)+\frac{r}{4N}+\frac{r\ln r-r-N\ln N + N +\ln \frac{N}{r}}{2N}-\frac{1}{q}
\\
&=\frac{1}{2}\left(1-\frac{r}{N}\right)(1+\ln N)+\frac{r}{4N}+\frac{r\ln r}{2N}-\frac{r}{2N}-\frac{\ln N}{2} + \frac{1}{2} +\frac{\ln N}{2N}
\nonumber
\\
& \phantom{aaaaaaaaaaaaa}-\frac{\ln r}{2N}-\frac{1}{q}
\\
&=\frac{1}{2}-\frac{r}{2N}+\frac{1}{2}\ln N-\frac{r}{2N}\ln N-\frac{r}{4N}+\frac{r-1}{2N}\ln r-\frac{1}{2}\ln N + \frac{1}{2} -\frac{1}{q}
\\
&=1-\frac{3r}{4N}-\frac{r}{2N}\ln N+\frac{r-1}{2N}\ln r-\frac{1}{q}
\\
&\approx 1-\frac{r-1}{2N}\left ( \frac{3}{2}+\ln \frac{N}{r}\right ) -\frac{1}{q} = 1-\frac{qm}{4N}\left ( \frac{3}{2}+\ln \frac{N}{1+\frac{qm}{2}}\right ) -\frac{1}{q},
\label{eq:RGGmod_copy}
\end{align}
from which we immediately recover Eq.(\ref{eq:RGGmod}), one of the main results of the paper.

\subsection{Expected modularity of different types of RGGs}
\label{subsec:scaling}

Let us assume that $\left<k(N)\right> = c_1 N^{\kappa}$ and $q(N) = c_2 N^{\lambda}$, where $c_1, c_2$ are constants, while $\lambda\in [0,1]$ and $\kappa\in [0,2]$ are scaling exponents. Substituting these scaling forms back to  Eq.(\ref{eq:RGGmod_copy0}) and Eq.(\ref{eq:RGGmod_copy}) we obtain
\begin{equation}
    \bar{Q}_{\text{RGG}}=\left\{\begin{matrix}
1-\frac{c_{1}c_{2}}{8}N^{\kappa + \lambda -1} \left ( \ln \frac{N}{1+\frac{c_1c_2}{4}N^{\kappa + \lambda}} +\frac{3}{2} \right ) -\frac{1}{q}, \ \ \ \text{if} \ \ \ r<N,
\\ 
\frac{N}{c_1c_2N^{\kappa + \lambda}}-\frac{1}{q}, \ \ \ \text{if} \ \ \ r\geq N,
\end{matrix}\right.  
\label{eq:RGGmod_scaling}
\end{equation}
where $qm\approx q\frac{\left<k\right>}{2}=\frac{c_1c_2}{2}N^{\kappa + \lambda}\sim N^{\kappa + \lambda}$  and $r=1+\frac{qm}{2}\approx 1+\frac{q\left<k\right>}{4}=1+2CN^{\kappa + \lambda}$.
Furthermore, introducing $C=c_1c_2$ and taking the asymptotic $N\to\infty$ limit, Eq.(\ref{eq:RGGmod}) can be further simplified to
\begin{equation}
    \bar{Q}_{\text{RGG}}\approx \left\{\begin{matrix}
1-C N^{\kappa + \lambda -1} \left ( \ln \frac{1}{2C} N^{1-(\kappa+\lambda)} +\frac{3}{2} \right ) -\frac{1}{c_2 N^{\lambda}}, \ \ \ \text{if} \ \ \ r<N,
\\ 
\frac{1}{8C} N^{1-(\kappa+\lambda)}-\frac{1}{q}, \ \ \ \text{if} \ \ \ r\geq N.
\end{matrix}\right.   
\end{equation}
Now, let us turn to the analysis of the various scenarios based on the values of $\lambda$ and $\kappa$. First, if $\lambda+\kappa>1$, then $r=1+2CN^{\lambda+\kappa}>N$ for large values of $N$, implying that Eq.(\ref{eq:RGGmod_scaling}) reduces to $\lim\limits_{N\to\infty}\bar{Q}_{\text{RGG}}=-\frac{1}{q}\sim-\frac{1}{N^{\lambda}}$. If the network is dense, that is $\kappa > 1$, then based on Eq.(\ref{eq:RGGmod_scaling}) we obtain $\lim\limits_{N\to \infty}\bar{Q}_{\text{RGG}}\sim -\frac{1}{N^{\lambda}}\to 0^{-}$, until assuming $\lambda\in (0,1]$. If $\lambda=0$ (along with $\kappa>1$), i.e. $q = c_2$, the expected modularity in Eq.(\ref{eq:RGGmod_scaling}) takes a negative constant value $\bar{Q}_{\text{RGG}}=-\frac{1}{c_2}$.  

Let us now turn to the case, where certain nodes $s$ connect, in expectation, to all other nodes in their community, meeting the condition
\begin{equation}
\bar{b}_s =\frac{N-1}{q}.
\end{equation} 
In this case, the expected modularity has an upper bound of $\bar{Q}_{\text{RGG}}=\frac{1}{4}$, which can be derived as follows
\begin{equation}
\bar{Q}_{\text{RGG}} = \frac{N}{2mq}-\frac{1}{q}=\frac{N-2m}{2mq}<\frac{N-2m}{4N-4}<\frac{N-1}{4N-4}<\frac{1}{4},
\end{equation}
where we used that $r=1+\frac{qm}{2}>N$, which implies $mq>2N-2$. Subsequently, let's examine the precise circumstances under which this phenomenon occurs. Initially, observe that if $s''>N$ in Eq.(\ref{eq:I2_global}) and $s'>s$ in Eq.(\ref{eq:I1_global}) both hold simultaneously, then $\Pi_q(s,t) = \frac{1}{q} \ \forall s,t$, leading to $\bar{b}_s=\frac{N-1}{q}$ based on Eq.(\ref{eq:QI1I2}). Furthermore, by unfolding the definition of $s''$ in Eq.(\ref{eq:t_doubleprime}) we obtain
\begin{align}
    s''&=\left ( 1+ \frac{qm}{2} \frac{1-\beta}{s^{\beta}}\right )^{1/(1-\beta)} > N
\label{eq:nec_condition}
\end{align}
which after rearrangement, can equivalently be written as
\begin{equation}
    \left( \frac{qm}{2} \frac{1-\beta}{N^{1-\beta}-1} \right )^{1/\beta}> s.
    \label{eq:sprimesdoubleprime}
\end{equation}
We can relate the inequality in Eq.(\ref{eq:sprimesdoubleprime}) to the parameter $s'$ defined in Eq.(\ref{eq:t_prime}), by observing that
\begin{equation}
    s'=\left ( \frac{qm}{2} \frac{1-\beta}{s^{1-\beta}-1}\right )^{1/\beta}>\left( \frac{qm}{2} \frac{1-\beta}{N^{1-\beta}-1} \right )^{1/\beta}> s.
\label{eq:nec_condition_conseq}
\end{equation}
The inequalities in Eqs.(\ref{eq:nec_condition}-\ref{eq:nec_condition_conseq}) collectively imply that if $s''>N$, then the condition $s'>s$ automatically holds as well. Consequently, the necessary condition for any arbitrary node $s$ to connect, on average, to all other nodes within its community is $s''>N$. Along this line, in order to meet the previous criterion for any node $s=1,...,N$, we require that
\begin{equation}
    N<\left ( \frac{qm}{2}\frac{1-\beta}{N^{1-\beta}-1} \right )^{1/\beta}
\end{equation}
which after rearrangement yields
\begin{equation}
  N <\frac{qm(1-\beta)}{2}
\end{equation}
for $\beta\in (0,1)$. Similar considerations suggest that the analogous necessary condition reads
\begin{equation}
    N\ln N <\frac{qm}{2}
\end{equation}
for $\beta = 1$.

As a next step, let us consider the case when $\lambda + \kappa = 1$. In this limit, the asymptotic modularity of RGG graphs are already very high. To see this, let us assume that $1+\frac{qm}{2}<N$, or equivalently $C\leq \frac{1}{2}-\frac{1}{2N}$, based on which, we obtain
\begin{equation}
        \bar{Q}_{\text{RGG}}\approx 1-C \left ( \ln \frac{1}{2C} +\frac{3}{2} \right ) -\frac{1}{c_2 N^{\lambda}}.
        \label{eq:1per4}
\end{equation}
Note that the expression in Eq.(\ref{eq:1per4}) is asymptotically minimized when $C$ is maximal. 
Using the fact that $\max \{ 
 \lim\limits_{N\to\infty}C\}=\frac{1}{2}$, 
the minimal asymptotic modularity in this specific limit can be estimated as
\begin{equation}
        \bar{Q}^{\text{min}}_{\text{RGG}}\approx 1-\frac{1}{2} \left ( \ln \frac{1}{2\frac{1}{2}} +\frac{3}{2} \right )=\frac{1}{4}.
        \label{eq:1per4_2}
\end{equation}
for $\lambda>0$, and 
\begin{equation}
        \bar{Q}^{\text{min}}_{\text{RGG}}\approx 1-\frac{1}{2} \left ( \ln \frac{1}{2\frac{1}{2}} +\frac{3}{2} \right )-\frac{1}{2}=-\frac{1}{4}.
        \label{eq:1per4_3}
\end{equation}
for $\lambda = 0$.

As mentioned in the main text of the article, when the sum of $\lambda + \kappa$ falls below the critical threshold of $1$ and $\lambda > 0$ holds, we arrive to a regime where Random Geometric Graphs display notably high modularity. In this highly modular scenario, networks must consistently maintain sparsity, indicating that $\kappa < 1$, and the number of communities $q$ should scale slower than $N^{1-\kappa}$. Until these conditions are satisfied, the two terms with negative pre-factors in Eq.(\ref{eq:RGGmod_scaling}) both converge to 0. Therefore, ultracold PSO networks with $\beta=0$ consistently achieve maximal modularity, that is
\begin{equation}
    \bar{Q}_{\text{RGG}}= 1.
\end{equation}
In cases when sparsity holds, but $\lambda=0$, RGG graphs are still characterized by high modularity values, albeit never maximal. This can be seen by setting $\lambda=0$ and $\kappa < 1$ in Eq.(\ref{eq:RGGmod_scaling}), which yields
\begin{equation}
\bar{Q}_{\text{RGG}}=1-\frac{1}{q},    
\end{equation}
from which we recover the result of Ref.~\cite{Balogh2023}.

\bibliographystyle{unsrt}







\end{document}